\documentclass[11pt,a4paper]{article}
\pdfoutput=1
\usepackage{jheppub}
\usepackage{slashed}

\makeatletter
\def\@fpheader{\relax}
\makeatother

\usepackage[czech,english]{babel}
\usepackage{graphicx}
\usepackage{amsmath,amsfonts,amssymb}
\usepackage{url}
\DeclareMathOperator{\MyProd}{\scalebox{1.4}{$\mathrm{I\kern-0.2ex I}$}}

\preprint{LCTP-19-15}

\title{AdS$_5$ Black Hole Entropy near the BPS Limit}

\author[a]{Finn Larsen}

\emailAdd{larsenf@umich.edu}

\author[a]{, Jun Nian}

\emailAdd{nian@umich.edu}

\affiliation[a]{Leinweber Center for Theoretical Physics, University of Michigan, Ann Arbor, MI 48109, U.S.A.}

\author[a]{and Yangwenxiao Zeng}

\emailAdd{zengywx@umich.edu}

\abstract{We analyze AdS$_5$ black holes that are nearly supersymmetric. They depart from the BPS limit in two distinct ways: a temperature takes them above extremality and a potential maintains extremality but violates a certain constraint. We study the thermodynamics of these deformations and their interplay in detail. We discuss recent microscopic computations of BPS black hole entropy in $\mathcal{N}=4$ SYM and 
generalize the arguments to the nearBPS regime by relaxing constraints imposed by supersymmetry. Our methods recover gravitational results from microscopic theory also for nearBPS black holes.}


\arxivnumber{}

\newcommand{\bea}{\begin{eqnarray}}
\newcommand{\eea}{\end{eqnarray}}

\newcommand{\be}{\begin{equation}}
\newcommand{\ee}{\end{equation}}

\begin{document}

\maketitle

\section{Introduction}\label{sec:introduction}

The microscopic understanding of black hole entropy is the linchpin for progress in quantum gravity. However, studies of black hole entropy have been quantitatively successful only in a few systems with a high degree of supersymmetry, e.g. \cite{Strominger:1996sh, Benini:2015eyy}. Such settings offer great control but they appear far removed from the studies of black hole dynamics that is focus of much current research, such as the study of the SYK model, e.g. \cite{Sachdev:1992fk, Kitaev,Maldacena:2016upp} or other fashionable spects of holography e.g. \cite{Gao:2016bin,Rangamani:2016dms,Harlow:2018fse}. The main motivation for this article is to develop an arena that has the potential to bridge this gap. We leverage results on BPS ground states obtained through in 4d $\mathcal{N}=4$ SYM to study the finite temperature properties of these systems. 

Supersymmetric black holes in AdS$_5$ offer a particularly important setting for the study of holography. The relevant classical geometries have been known for quite some time \cite{Gutowski:2004ez, Gutowski:2004yv, Chong:2005hr, Wu:2011gq}. It is an unfortunate technical complication that these black hole solutions are necessarily somewhat complicated. For example, all supersymmetric AdS$_5$ black holes with regular event horizon rotate. On the other hand, the entropy of these black holes is relatively simple \cite{Kim:2006he}:
\be\label{eq:Entropy}
  S = 2 \pi \sqrt{Q_1 Q_2 + Q_2 Q_3 + Q_3 Q_1 - \frac{1}{2} N^2(J_1 + J_2)}\, ,
\ee
where $Q_I$ (with $I=1,2,3$) denote the R-charges and $J_i$ (with $i=1,2$) the angular momenta within AdS$_5$. The charges of supersymmetric AdS black holes must satisfy not only a conventional
BPS mass condition 
\be
  M = \sum_{I=1}^3 Q_I + \sum_{i=1}^2 J_i\, ,
\ee
but also a certain nonlinear constraint.
\be\label{eq:Constraint}
  Q_1 Q_2 Q_3 + \frac{1}{2}N^2 J_1 J_2 = \left(\frac{1}{2}N^2 + Q_1 + Q_2 + Q_3 \right) \left(Q_1 Q_2 + Q_2 Q_3 + Q_3 Q_1 - \frac{1}{2}N^2 (J_1 + J_2) \right)\, .
\ee
The full significance of the constraint is somewhat mysterious. On the gravity side, some researchers impose it as the condition that closed timelike curves are absent. We will trace it to a simpler regularity condition.

It was long thought that supersymmetric indices fail at counting the BPS states underlying the entropy of supersymmetric AdS$_5$ black holes. For example, 
all possible indices were constructed~\cite{Kinney:2005ej} and their growth was estimated as $\mathcal{O} (1)$. This result was interpreted as due to large cancellations in the partition function that preclude the growth of $\mathcal{O} (N^2)$ that is needed to account for the black hole entropy. 
Further work included \cite{Berkooz:2006wc,Grant:2008sk,Chang:2013fba}.

There has been progress on this research over the last few years. A central insight was the recasting of the entropy \eqref{eq:Entropy} and the constraint \eqref{eq:Constraint} in terms of the free energy \cite{Hosseini:2017mds}:
\be\label{eq:FreeEnergy}
  \ln \, Z = -  \frac{1}{2} N^2 \frac{\Delta_1 \Delta_2 \Delta_3}{\omega_1 \omega_2}\, ,
\ee
with the potentials satisfying a complex constraint. 
A straightforward Legendre transform of this expression from the potentials for R-charge $\Delta_I$ and the rotational velocities $\omega_i$ gives the entropy \eqref{eq:Entropy}. Moreover, the constraint \eqref{eq:Constraint} on charges follows in the process, from a reality condition on the entropy. Subsequently, the free energy and its accompanying constraint were derived from the on-shell action of supergravity \cite{Cabo-Bizet:2018ehj}. These results (and their generalizations to 
other dimensions) appear deeply interrelated with the supersymmetric 
Casimir energy \cite{Assel:2014paa,Assel:2015nca} and its relation to anomalies \cite{Bobev:2015kza,Brunner:2016nyk}.

Over the last few months several microscopic derivations of the BPS entropy of AdS$_5$ black holes have been presented:
\begin{itemize}
\item
{\it Supersymmetric Localization} \cite{Cabo-Bizet:2018ehj}. This ab initio computation is principled: the central point is the deformation of the path integral away from the physical surface while preserving fermion boundary conditions consistent with supersymmetry. However, the argument relies on a somewhat mysterious ``generalized supersymmetric Casimir energy'' that relates the supersymmetric partition function (which is of order $\mathcal{O} (N^2)$) and the supersymmetric index (which is of order $\mathcal{O} (1)$). 

\item
{\it Free Field Construction} \cite{Choi:2018hmj}. This is the simplest derivation by far, involving nothing but the free field representation of $\mathcal{N}=4$ SYM at vanishing coupling. However, it is not clear to us that the computation is justified in the regime where it is applied. 

\item
{\it Superconformal Index} \cite{Benini:2018ywd}. This computation relies on the earlier rewriting of the $\mathcal{N}=4$ SYM localization on $S^1 \times S^3$ in terms of a complex integral over a circle, with poles inside the disc determined by certain Bethe Ansatz equations \cite{Benini:2018mlo}. Manipulating the contour and taking advantage of known poles of the integrand \cite{Hosseini:2016cyf,Hong:2018viz}, the free energy \eqref{eq:FreeEnergy} is extracted from the supersymmetric index. This derivation is the most rigorous but it is technically more involved and was so far completed only for simplified values of the angular momenta.
\end{itemize}
A key technical feature of all these derivations is that they invoke complexified potentials $\Delta_I$, $\omega_i$ in an essential manner. However, it is not obvious to us that the details of these papers are consistent with one another \footnote{For some discussions see \cite{Cabo-Bizet:2018ehj,Zaffaroni:2019dhb}.}. Further research prompted by \cite{Cabo-Bizet:2018ehj, Choi:2018hmj, Benini:2018ywd} has generalized details of the results in various directions \cite{Choi:2018vbz, Honda:2019cio, ArabiArdehali:2019tdm, Choi:2019miv, Kim:2019yrz, Cabo-Bizet:2019osg, Amariti:2019mgp}
but no consensus has emerged yet.

\subsection*{Summary of this Article}

In this paper we take the substantiation of the BPS free energy (\ref{eq:FreeEnergy}) for granted. Given that basis, we make a significant leap and study the entropy of AdS$_5$ black holes {\it away from the supersymmetric limit}. We develop both gravitational and microscopic considerations, and we find several new agreements between these holographically dual descriptions. Our success in this direction develops the emerging microscopics of AdS$_5$ black holes directly in the regime that is physically relevant. 

It is important to recognize that AdS$_5$ black holes allow {\it two distinct deformations} away from the BPS limit. Conceptually, an extremal limit $M=M_{\rm ext}$ indicates the lowest possible mass for given conserved charges. The low lying excitations with energy $M$ in the range $0<M - M_{\rm ext} \ll M_{\rm ext}$ are characterized quantitatively by the low temperature behavior
\be
\label{eqn:CTTfirst}
  M - M_{\rm ext} = \frac{1}{2}\, \left(\frac{C_T}{T}\right)\, T^2\, ,
\ee
where $C_T$ is the specific heat.\footnote{The specific heat $C_T$ at low temperature is proportional to the temperature so the combination $\frac{C_T}{T}$ is a constant in the regime we study. In our notation $\frac{C_T}{T}$ is the coefficient in front of ${1\over 2}T^2$, evaluated with all charges $Q_I$, $J_i$ kept fixed. Some might refer to this variable as $C_{Q, J}$ but that is not what we do.} The regime where this mass formula applies is also studied in research inspired by the SYK model \cite{Sachdev:1992fk, Kitaev}, such as \cite{Almheiri:2014cka, Maldacena:2016upp, Almheiri:2016fws, Larsen:2018iou}.

However, in the context of AdS$_5$ black holes, it is equally possible to consider excitations away from the BPS limit that remain on the extremal surface $M = M_{\rm ext}$  for given conserved charges $Q_I$, $J_i$ but with those charges taking values that violate the constraint \eqref{eq:Constraint}. We find that such excitations have mass 
\be
\label{eqn:cvarphidef}
  M - M_{\rm BPS} = \frac{1}{2}\, \left(\frac{C_\varphi}{T}\right)\, \left(\frac{\varphi}{2 \pi\ell_5}\right)^2\, , 
\ee
where $\varphi$ is a potential parametrizing departures from the BPS surface that preserve extremality and so have $T=0$.  The coefficient $C_\varphi$ is the {\it capacitance} of the 
black hole. \footnote{It is admittedly ${C_\varphi\over T}$ that is the capacitance according to the definition \eqref{eqn:CTTfirst}. We find this abuse of language an acceptable price for making 
the symmetry between $C_T$ and $C_\varphi$ manifest.}
We give the precise relation between the deformations $\varphi$ within the extremal surface and the generic potentials $\Delta_I$, $\omega_i$ in \eqref{eqn:defconstra} with $T=0$. 
To complete the set of parameters describing linear response we also introduce a thermoelectric coefficient $C_E$ that quantifies the interplay between the temperature $T$ and the potential $\varphi$. 

The computation of the response parameters $C_T$, $C_\varphi$, and $C_E$ is conceptually straightforward on the supergravity side, albeit not trivial from a technical point of view. 
Surprisingly, we find that the heat capacity and the capacitance are identical
\be
  C_T = C_\varphi. 
\ee
Each of these physical quantities are quite nontrivial functions of black hole parameters. They are not related in any obvious matter in supergravity and any symmetry between them would be novel and interesting in ${\cal N}=4$ SYM. 

On the microscopic side any progress may seem implausible because advances in the BPS limit rely heavily on supersymmetry, not to mention that they face some unresolved questions. 
However, the nonrenormalization due to supersymmetry may generalize to the linear order that we study. Certainly the low temperature heat capacity is subject to a nAttractor 
mechanism \cite{Larsen:2018iou}. Independently, the mass terms that depend quadratically on potentials were derived from BPS considerations in the case of asymptotically flat black 
holes \cite{Larsen:1999uk,Larsen:1999dh}. It is therefore reasonable to expect that the parameters we compute are protected, even though we have not worked out a detailed argument to this effect. 

In our microscopic computations we proceed pragmatically. The BPS free energy (\ref{eq:FreeEnergy}) only gives the limit of the near extremal partition function but smoothness is sufficient to find the linear dependence on temperature even away from the BPS limit. Moreover, our gravitational study motivate relaxing the constraint imposed by supersymmetry as well. It is satisfying that the two independent deformations combine nicely, to a complex parameter $\varphi + 2\pi i T$. With these minimal and conservative ingredients we recover gravitational results through a Legendre transform.

This paper is organized as follows. In section~\ref{sec:BHThermo} we develop the gravitational thermodynamics of nearBPS AdS$_5$ black holes. We carefully distinguish between the near-extremal limit and the extremal nearBPS limit, and we study their interplay. In section~\ref{sec:FreeFT} we review the partition function of $\mathcal{N}=4$ SYM in the free field limit, which can be written as a matrix model. 
We derive the free energy (\ref{eq:FreeEnergy}) from its leading contribution in the large-$N$ and low-temperature limit. In section~\ref{sec:BHStatMech} we study the resulting AdS$_5$ black hole entropy function. We derive some of its implications beyond the regime where it was originally derived, by relaxing constraints on potentials and exploiting general principles such as the first law of thermodynamics. This leads to
statistical physics of nearBPS AdS$_5$ black holes that agrees with results from the gravity side. Some open questions and future directions are discussed in section~\ref{sec:discussion}.

\section{Black Hole Thermodynamics}\label{sec:BHThermo}
 
In this section, we introduce the thermodynamics of AdS$_5$ nearBPS black holes. We consider the general AdS$_5$ black hole before taking the BPS limit, and then study deviations from the BPS limit by changing temperature or physical charges. We consider the near-extremal limit and the extremal nearBPS limit in subsections \ref{sec: near_ext} and \ref{sec: ext near BPS}, respectively, and then generalize our results to general infinitesimal deviations from the BPS surface in subsection~\ref{sec: near BPS}.
    
The general AdS$_5$ black hole solution (for any $M$, $Q_I$, $J_i$) is known \cite{Wu:2011gq} but has not been analyzed in much detail \cite{Birkandan:2014vga}. In this section we follow most of the literature and focus on diagonal R-charges $Q_1 = Q_2 = Q_3$.

\subsection{General AdS$_5$ Black Holes}\label{sec: General BH solutions}
We consider the most general charged rotating black holes in five-dimensional minimal gauged supergravity \cite{Chong:2005hr}. The solutions are identified by the mass $M$, the charge $Q$, and two independent angular momenta $J_i$ (with $i=1,2$). These physical quantum numbers are parametrized by four variables $(m,q,a,b)$ as
    \begin{align}
        M & = \frac{\pi}{4G_5}\frac{m(2\Xi_a+2\Xi_b-\Xi_a\Xi_b)+2qabg^2(\Xi_a+\Xi_b)}{\Xi_a^2\Xi_b^2}~,\label{eqn: mass}\\
        Q & = \frac{\pi}{4G_5}\frac{q}{\Xi_a \Xi_b}~,\label{eqn: charge}\\
        J_1 & = \frac{\pi}{4G_5}\frac{2ma+qb(1+a^2g^2)}{\Xi_a^2 \Xi_b}~,\\
        J_2 & = \frac{\pi}{4G_5}\frac{2mb+qa(1+b^2g^2)}{\Xi_a \Xi_b^2}~. \label{eqn: angular momenta}
    \end{align}
Here $G_5$ is Newton's gravitational constant in five dimensions, $g=\ell_5^{-1}$ with $\ell_5$ the AdS$_5$ radius, and
\be
  \Xi_a \equiv 1 - a^2 g^2\, ,\quad \Xi_b \equiv 1 - b^2 g^2\, .
\ee

The coordinate that locates the event horizon $r_+$ is the largest real root of $\Delta_r=0$, where $\Delta_r$ is given by
\begin{eqnarray}\label{eqn: horizon}
\label{Delta_r}&&\Delta_r=\frac{(r^2+a^2)(r^2+b^2)(1+g^2r^2)+q^2+2abq}{r^2}-2m~.
\end{eqnarray}
We can expresses $m$ in terms of $r_+^2$ and other variables using this equation. Then the temperature and entropy are conveniently expressed in terms of $(r_+,q,a,b)$ as
\begin{eqnarray}
\label{Temperature}&&T=\frac{r_+^4[1+g^2(2r_+^2+a^2+b^2)]-(ab+q)^2}{2\pi r_+[(r_+^2+a^2)(r_+^2+b^2)+abq]}~,\\
\label{eqn: entropy}&&S= 2 \pi \frac{\pi}{4 G_5}\frac{(r_+^2+a^2)(r_+^2+b^2)+abq}{(1-a^2g^2)(1-b^2g^2)r_+}~.
\end{eqnarray}
The electric potential and angular velocities on the horizon are given in the same notation by
\begin{align}
\begin{split}\label{eqn: potential}
\Phi & = \frac{3qr_+^2}{(r_+^2+a^2)(r_+^2+b^2)+abq}~,\\
\Omega_1 & = \frac{a(r_+^2+b^2)(1+g^2r_+^2)+bq}{(r_+^2+a^2)(r_+^2+b^2)+abq}~,\\
\Omega_2 & = \frac{b(r_+^2+a^2)(1+g^2r_+^2)+aq}{(r_+^2+a^2)(r_+^2+b^2)+abq}~.
\end{split}
\end{align}

\subsection{BPS AdS$_5$ Black Holes}\label{sec: BPS BH solutions}


The general black holes introduced in the previous subsection depend on independent physical variables $(M, Q, J_1, J_2)$. Supersymmetry of the theory guarantees that their mass satisfies 
$$
M - \left( 3 Q + g J_1 + gJ_2)\right)  \geq 0~.
$$
The BPS black holes saturate this inequality so their mass $M$ is given by the BPS condition
\begin{equation}\label{eqn:mqjbps}
M^* = 3 Q^* + g (J_1^* + J_2^*)~.
\end{equation}
We use the starred symbols $(M^*, Q^*, J_1^*, J_2^*)$ instead of $(M, Q, J_1, J_2)$ when we stress that the variables refer to the BPS case. \footnote{The star must not be confused with complex conjugation.}

The general formula for the mass can be written as 
\begin{equation}
\label{eqn:Mmrelation}
M - \left( 3 Q + g J_1 + gJ_2\right)  = {\pi\over 4G_5} {3 + (a+b)g - abg^2\over (1-ag)(1+ag)^2(1-bg)(1+bg)^2} \left[ m - q(1+g(a+b)\right]~,
\end{equation}
in terms of the parameters $(m, q, a, b)$. The coefficient that depends on $(a,b)$ is always positive, so the physical BPS condition \eqref{eqn:mqjbps} is equivalent to the relation
\begin{eqnarray}\label{BPS condition}
&&q^*=\frac{m^*}{1+ag+bg}~.
\end{eqnarray}
It turns out that once we impose the BPS condition, regularity of the underlying black hole geometry further imposes a constraint between the black hole charges. This constraint takes the form 
\begin{eqnarray}\label{constraint}
&&q^*=\frac{1}{g}(a+b)(1+ag)(1+bg)~, 
\end{eqnarray}
when expressed in terms of the parameters $(q, a, b)$. The two conditions (\ref{BPS condition}-\ref{constraint}) that must be satisfied by BPS black holes together yield starred variables $(m^*,q^*)$ that are 
specified functions of $(a,b)$. 

It is clearly convenient to pick the ``rotation" parameters $(a, b)$ as the two independent coordinates on the BPS surface. We can express the physical variables as:
\begin{align}
M^* & = \frac{\pi}{4G_5}\frac{\left(3(a+b)-(a^3+b^3) g^2-ab(a+b)^2 g^3\right)}{g (1-a g)^2 (1-b g)^2}~,\label{eqn:BPS Mass}\\
Q^* & = \frac{\pi}{4G_5} \frac{a+b}{g (1-a g) (1-b g)}~,\label{eqn:BPS Charges}\\
 J_1^* & = \frac{\pi}{4G_5}\frac{(a+b) (2a +b+abg)}{g (1-a g)^2 (1-b g)}~,\label{eqn:BPS Ja}\\
 J_2^* & = \frac{\pi}{4G_5}\frac{(a+b) (a+2b+abg)}{g (1-a g) (1-b g)^2}~.\label{eqn:BPS Jb}
\end{align}
These expressions satisfy the BPS condition (\ref{eqn:mqjbps}) for any $(a,b)$, as they must.

For BPS black holes the mass $M^*$ is never an independent parameter, and in our setting it is the linear function \eqref{eqn:mqjbps} of the other charges. It is a less familiar feature that the condition \eqref{constraint} imposes an additional relation. This is the reason that the three variables $Q^*, J^*_{1,2}$ (\ref{eqn:BPS Charges} - \ref{eqn:BPS Jb}) are expressed in terms of just two coordinates $(a, b)$ on the BPS surface. It means these charges are not independent in the BPS case, they satisfy the {\it constraint}: 
\begin{align}\label{eqn: Charge Constraint}
 Q^{*3} + \frac{\pi}{4G_5} J_1^* J_2^* = \left(\frac{\pi}{4g^2 G_5}+3Q^* \right)\left(3Q^{*2}-\frac{\pi}{4gG_5}(J_1^* + J_2^*)\right)\, .
\end{align}
This is the special case of the constraint \eqref{eq:Constraint} with diagonal R-charges $Q_1 = Q_2 = Q_3$.

The general formulae \eqref{eqn: potential} for the electric potential and the angular velocities are nearly trivial in the BPS limit. They give $\Phi^* = 3$ and $\Omega_1^* = \Omega_2^* =g$ for all BPS black holes, independent of the values of $(a, b)$. These values are guaranteed by the first law of thermodynamics
\be\label{eq:FirstLawBPS}
  T^* dS^* = dM^* - \Phi^* dQ^* - \Omega_1^* dJ_1^* - \Omega_2^* dJ_2^*~,
\ee
because in the BPS case $T^*=0$ and the mass $M=M^*$ is given by \eqref{eqn:mqjbps}.

When the BPS conditions (\ref{BPS condition}-\ref{constraint}) are imposed, the largest root of the horizon equation $\Delta_r=0$ where $\Delta_r$ is given in \eqref{eqn: horizon} is a double root. This situation corresponds
to temperature $T=0$ and is expected for any extremal black hole. The double root has the simple value:
\begin{eqnarray}\label{r_+}
&&r^* \equiv r_+=\sqrt{\frac{1}{g}(a+b+abg)}~.
\end{eqnarray}
Given this value for $r_+$ and the BPS value for $q^*$ given in \eqref{BPS condition}, the general formula \eqref{eqn: entropy} gives the black hole entropy entirely in terms of the rotation parameters $(a, b)$:
\begin{eqnarray}\label{eqn: BPS entropy-1}
S^* = 2 \pi \cdot \frac{\pi}{4 G_5}\frac{a+b}{g (1-a g) (1-bg)}\sqrt{\frac{1}{g}(a+b+abg)} \, .
\end{eqnarray}

In our manipulations we will often need the inverse of (\ref{eqn:BPS Charges}-\ref{eqn:BPS Jb}) and so express $(a,b)$ in terms of the physical variables $Q^*$ and $J_{1, 2}^*$ of the BPS black hole. The resulting formulae are not unique, because the BPS charges are subject to the constraint \eqref{eqn: Charge Constraint}. A simple version is
\begin{align}
  a & = \frac{ 2 g Q^{*2} - {\pi\over 4G_5} J_2^*}{Q^* ( 2 g^2 Q^* + {\pi\over 4G_5})}\, ,\label{eqn:a and charges}\\
  b & = \frac{ 2 g Q^{*2} - {\pi\over 4G_5} J_1^*}{Q^* ( 2 g^2 Q^* + {\pi\over 4G_5})}\, .\label{eqn:b and charges}
\end{align}
As an example, we can use these equations and the constraint \eqref{eqn: Charge Constraint} to recast the BPS entropy \eqref{eqn: BPS entropy-1} as \cite{Kim:2006he}
\begin{eqnarray}\label{eqn: BPS entropyn}
S^* = 2\pi\sqrt{3(\ell_5  Q^*)^2-{\pi\over 4G_5} \ell^3_5 \left(J_1^* + J_2^* \right)}~.
\end{eqnarray}

In this section we have so far made the effort to retain the dimensionful scales $G_5$ and $g = \ell_5^{-1}$ in all our equations. This is common in supergravity equations that are written in terms of parametric variables $(m, q, a, b)$, and not terribly inconvenient. However, the practice becomes cumbersome when rewriting formulae in terms of conserved charges $M^*$, $Q^*$, and $J_{1, 2}^*$. It is better to employ the dimensionless quantity
\be\label{eq:ParamRel}
  \frac{1}{2} N^2 = \frac{\pi }{4 G_5} ~ \ell_5^3\, ,
\ee
where $N$ is the rank of the dual $SU(N)$ gauge group. For example, it is superior to express the BPS entropy \eqref{eqn: BPS entropyn} as 
\begin{eqnarray}\label{eqn: BPS entropy}
S^* = 2\pi\sqrt{3(\ell_5  Q^*)^2-\frac{1}{2} N^2 \left(J_1^* + J_2^* \right)}~.
\end{eqnarray}
To compare with microscopic results we additionally record that the supergravity charge $Q$ (with dimension $\textrm{length}^{-1}$) and the (dimensionless) angular momenta $J_{1, 2}$ are normalized such that $Q\ell_5$ and $J_{1,2}$ can be identified with the quantized charges in the microscopic theory. For clarity, we will retain the scale $g = \ell_5^{-1}$ explicitly in the remainder of this section, but these units will be dropped later in the article.

  \subsection{Near-Extremal Limit}\label{sec: near_ext}
  
Conceptually, the simplest way to deform away from the BPS surface is by adding energy to the BPS black hole, while {\it keeping electric charge and angular momenta fixed}. Such deviations take the black hole away from extremality and lead to nonzero temperature. This is the situation we consider in this subsection.
 
Since we consider charges that are fixed at their BPS values $Q^*$, $J_{1, 2}^*$ we can write the first law of thermodynamics in the near-extremal limit as
\be\label{eq:FirstLawNearExt}
  TdS = dM - \Phi dQ^* - \Omega_1 dJ_1^* - \Omega_2 dJ_2^*\, .
\ee
Subtracting the corresponding BPS expression \eqref{eq:FirstLawBPS} we have
\be\label{eq:FirstLawDiff-NearExt}
T dS = d (M - M^*) - (\Phi - \Phi^*) dQ^* - (\Omega_1 - \Omega_1^*) dJ_1^* - (\Omega_2 - \Omega_2^*) dJ_2^*\, .
\ee
Variations {\it along} the BPS surface have $M=M^*$ identically and correspond to the limit $T\to 0$ so they are described by 
\be
\label{eqn:BPSfirstlaw}
  dS = - (\partial_T \Phi) dQ^* - (\partial_T \Omega_1) dJ_1^* - (\partial_T \Omega_2) dJ_2^*\, . 
\ee
This formula describes the dependence on charges of the BPS entropy \eqref{eqn: BPS entropy}. That is interesting but not our focus in this subsection. 

Instead, we keep the charges strictly fixed and consider the heat added to the black hole as we raise the temperature from $T=0$ \cite{Preskill:1991tb}. 
After dividing by the small temperature $T$, we have 
\be\label{eq:NearExtdS}
  dS=\left(\frac{\partial S}{\partial T}\right)_{Q,J_{1,2}}dT=\frac{C_T}{T}\bigg|_{\text{nearExt}}~dT\, , 
\ee
where $C_T$ is the heat capacity that was introduced already in \eqref{eqn:CTTfirst}, along with comments on our notation. The 
ratio $\frac{C_T}{T}$ is constant in the near-extremal regime so:
\begin{eqnarray}
dM=TdS=\frac{1}{2}\frac{C_T}{T}\bigg|_{\text{nearExt}}~d\left(T^2\right)~.
\end{eqnarray}
After integration we have the leading behavior at small temperature
\begin{eqnarray}\label{eqn:M_Tsquare}
M-M^*=\frac{1}{2}\frac{C_T}{T}\bigg|_{\text{nearExt}}~T^2~,
\end{eqnarray}
where $M^* = 3 Q^* + g (J_1^* + J_2^*)$ is our reference point on the BPS surface. 
    
We can compute $C_T$ explicitly from the definition \eqref{eqn:BPSfirstlaw}, by brute force. We first use the general entropy formula \eqref{eqn: entropy} to evaluate the dependence of the entropy on black hole parameters through $\frac{\partial S}{\partial (r_+,q,a,b)}$.
We similarly compute the entries of the matrix $\frac{\partial (T,Q,J_1,J_2)}{\partial (r_+,q,a,b)}$, from the general formulae for temperature \eqref{Temperature} and charges (\ref{eqn: charge}-\ref{eqn: angular momenta}),
with the parameter $m$ eliminated in terms of $(r_+,q,a,b)$ using the horizon equation \eqref{eqn: horizon}. Inversion of this matrix (using Mathematica) gives the Jacobian $\frac{\partial (r_+,q,a,b)}{\partial (T,Q,J_1,J_2)}$ and then we can form
\begin{eqnarray}\label{heat capacity}
\frac{C_T}{T}\bigg|_{\text{nearExt}}  &=&\left(\frac{\partial S}{\partial r_+}\right)_{q,a,b}\left(\frac{\partial r_+}{\partial T}\right)_{Q,J_{1,2}}+\left(\frac{\partial S}{\partial q}\right)_{r_+,a,b}\left(\frac{\partial q}{\partial T}\right)_{Q,J_{1,2}}  \cr
&&+\left(\frac{\partial S}{\partial a}\right)_{r_+,q,b}\left(\frac{\partial a}{\partial T}\right)_{Q,J_{1,2}} +\left(\frac{\partial S}{\partial b}\right)_{r_+,q,a}\left(\frac{\partial b}{\partial T}\right)_{Q,J_{1,2}}\cr
&=&\frac{\pi}{4G_5}\frac{\pi^2 (a+b)^2 (3+(a+b)g-abg^2)}{g^2 (1-a g) (1-b g) \left(1+3(a+b)g+(a^2+3ab+b^2)g^2\right)}~.\label{eq:NearExtCT/T}
\end{eqnarray}
In the final step we exploited \eqref{constraint} and \eqref{r_+} to eliminate $r_+$ and $q$.
    
The heat capacity can be expressed as a function of BPS physical charges $Q^*$ and $J_{1,2}^*$. We first rewrite $(a,b)$ using (\ref{eqn:a and charges}-\ref{eqn:b and charges}) and then simplify 
using the constraint \eqref{eqn: Charge Constraint} between charges. The result is not unique, because of the constraint, but we find the manageable expression
\begin{eqnarray}\label{heat capacity in terms of charges}
         \frac{C_{T}}{T \ell_5}\bigg|_{\text{nearExt}}         
         &=&\pi^2 \frac{8 (Q^*\ell_5)^3 + \frac{1}{4} N^4 (J_1^* + J_2^*) }{3 (Q^*\ell_5 )^2-\frac{1}{2} N^2 \left(J_1^* + J_2^* \right) + \left(3 Q^*\ell_5  + \frac{1}{2} N^2\right)^2}~.
\end{eqnarray}
We can validate this result by inserting the formulae (\ref{eqn:BPS Charges} - \ref{eqn:BPS Jb}) for $Q^*$ and $J_{1,2}^*$ and restore units using \eqref{eq:ParamRel}. 
It is this form of the heat capacity that we can compare with microscopic considerations.

There is an alternative computation that leads to the heat capacity \eqref{eq:NearExtCT/T} with less effort and more insight. It is known as the nAttractor mechanism \cite{Larsen:2018iou}. 
The key observation is that for fixed conserved charge the derivative with respect to temperature $T$ and the horizon coordinate $r_+$ are equivalent. Therefore, it is sufficient to consider the effect caused by the change of $r_+$ for computation of the heat capacity. Moreover, the departure from the BPS mass $M-M^*$ given in \eqref{eqn:M_Tsquare} is quadratic in the temperature while the entropy is only linear. Therefore, at the linear order, it is sufficient to consider the BPS geometry, there is no need for the general black hole solution. This leads to the economical computation
\begin{eqnarray}
&&\frac{C_T}{T}\bigg|_{\text{nearExt}}=\left(\frac{\partial S}{\partial T}\right)_{Q,J_{1,2}}=\left(\frac{\partial S}{\partial r_+}\right)_{q,a,b}\left(\frac{\partial T}{\partial r_+}\right)^{-1}_{q,a,b}~.
\end{eqnarray}
This expression can be evaluated by hand in a few lines and gives the same result as \eqref{heat capacity}.

It is similarly useful to think of the electric potentials $\Phi$ and rotational velocities $\Omega_i$ as radially dependent ``attractor flows'' that take their fixed values $\Phi^*=3$, $\Omega_1^* = \Omega_2^* = g$ at the horizon. The final approach to the horizon is determined in each case by a radial derivative along the flow. For the electric potential we have
\begin{align}
 \left(\frac{\partial \Phi}{\partial T}\right)_{Q,J_{1,2}} & = \left(\frac{\partial \Phi}{\partial r_+}\right)_{q,a,b}\left(\frac{\partial T}{\partial r_+}\right)^{-1}_{q,a,b}\nonumber\\
{} & = -\frac{3 \pi  (a+b) \left(1-a b g^2\right)}{g \sqrt{\frac{a
b g+a+b}{g}} \left(1+3(a+b)g+\left(a^2+3 a b+b^2\right)g^2 \right)} \nonumber\\
{} & = -\frac{\pi^2 N^2 \ell_5}{S^*}~\frac{\left(J_1+J_2\right)\left(3 Q \ell_5+\frac{1}{2} N^2\right)-2\left(\frac{S^*}{2\pi}\right)^2}{\left(\frac{S^*}{2 \pi} \right)^2 + \left(3 Q \ell_5 + \frac{1}{2} N^2\right)^2}~. \label{eq:PhiadT}
\end{align}
In the final formula we used the BPS entropy $S^*$ given in \eqref{eqn: BPS entropy} as a preferred combination of charges, in order to avoid an expression that is overly unwieldy. 
   
For the temperature dependence of the rotational velocity we similarly find
\begin{align}
\left(\frac{\partial \Omega_1}{\partial T}\right)_{Q,J_{1,2}} & = \left(\frac{\partial \Omega_1}{\partial r_+}\right)_{q,a,b}\left(\frac{\partial T}{\partial r_+}\right)^{-1}_{q,a,b} \nonumber\\
{} & = -\frac{\pi  (1-a g) (a+2 b+(2 a+b)b g )}{\sqrt{\frac{a b
g+a+b}{g}} \left(1+3(a+b)g+\left(a^2+3 a b+b^2\right)g^2 \right)}~\nonumber\\
{} & = -\frac{\pi^2 N^2 \ell_5}{S^*}~\frac{J_2\left(3 Q \ell_5 + \frac{1}{2} N^2\right)-\left(\frac{S^*}{2\pi}\right)^2}{\left(\frac{S^*}{2 \pi} \right)^2 + \left(3 Q \ell_5 + \frac{1}{2} N^2\right)^2}~.\label{eq:dOmegaadT}
\end{align}
There is an analogous formula for $\partial_T\Omega_2$, given by exchanging $a\leftrightarrow b$ (or $J_1\leftrightarrow J_2$). 

The temperature dependence of the potentials given in (\ref{eq:PhiadT}-\ref{eq:dOmegaadT}) is such that the BPS limit of the first law \eqref{eqn:BPSfirstlaw} is satisfied. It is also interesting that
\begin{equation}
\label{eqn:potTdiff}
\left(\frac{\partial \Phi}{\partial T}\right)_{Q,J_{1,2}}  = \left(\frac{\partial \Omega_1}{\partial T}\right)_{Q,J_{1,2}} + \left(\frac{\partial \Omega_2}{\partial T}\right)_{Q,J_{1,2}}~.
\end{equation}
We will see shortly that this is a consequence of temperature {\it respecting} the constraint on charges \eqref{eqn: Charge Constraint}.
    
\subsection{Extremal NearBPS Limit}\label{sec: ext near BPS}

In this subsection we consider departures from the BPS surface that preserve extremality $T=0$. This situation is somewhat unusual. The nearBPS black holes we study remain extremal in the conventional sense of the black hole attaining its minimal possible mass for given charges. However, charges are modified such that the constraint \eqref{eqn: Charge Constraint} is violated. We recall that preserved supersymmetry requires BPS saturation which in turn implies the constraint on charges. We therefore conclude that the mass must exceed the BPS bound \eqref{eqn:mqjbps} and that the black holes do not preserve supersymmetry. 
    
The most important challenge will be to understand the extremal surface in more detail. According to \eqref{eqn:Mmrelation}, the mass $M$ generally exceeds the BPS mass $M^*$ by an amount that is proportional to
\begin{align}\label{eq:horizon recast}
{} & m - (1+ag+bg) q \nonumber\\
= & \, \frac{g^2 r_+^2 (q - q^*)^2+ \Big( \left[ (1 + a g + b g)^2 + g^2 r_+^2 \right] (r_+^2 - r^{*2}) - (1 + a g + b g) (q - q^*) \Big)^2}{2\, r_+^2 \big[ (1 + a g + b g)^2 + g^2 r_+^2 \big]}  \, ,
\end{align}
after some rewriting of the expression for $m$ that follows from the horizon equation $\Delta_r=0$ with $\Delta_r$ given in \eqref{eqn: horizon}. We recall that the starred variables $q^*$ and $r^*$ refer to the functions (\ref{constraint}, \ref{r_+}) of $(a,b)$ that apply on the BPS surface. The right hand side of \eqref{eq:horizon recast} is manifestly positive for real parameters so we recover the BPS bound in the form
\be
  m \geq (1 + a g + b g ) q\, ,
\ee
with equality on the BPS surface if and only if $q = q^*$ and $r_+ = r^*$. Therefore, the constraint on charges \eqref{eqn: Charge Constraint} follows from BPS saturation, neither an independent assumption nor a physical requirement is needed. 

However, the extremal surface is characterized by vanishing temperature, not by the BPS condition. Near the BPS surface we can take
\be\label{eq:ExtNearBPSVar}
  r_+^2 - r^{*2} \sim q - q^* \sim \epsilon\, ,
\ee
small and approximate the temperature $T$ \eqref{Temperature} as
\be
\label{eqn:Tapx}
  T = \frac{ \Big[1 + 3(a+b)g + (a^2 + b^2 + 3ab)g^2 \Big] (r_+^2 - r^{*2}) -  (1 + (a+b)g) \, (q - q^*)}{\pi r^* q^*}\, .
\ee
Thus extremality corresponds to a correlation between the magnitudes of $r_+^2 - r^{*2}$ and $q - q^*$ such that the two terms in the temperature \eqref{eqn:Tapx} cancel at linear order. 
In this regime the second term in the numerator of \eqref{eq:horizon recast} vanishes, since it is proportional to the temperature, but the first term does not. Thus the BPS condition is preserved at linear order in $\epsilon$, but it is broken at quadratic order. This structure is reminiscent of the nAttractor arguments reviewed in the preceding subsection, but with departures from the BPS surface now due to charges that violate the constraint, rather than nonzero temperature. We also learn 
that by taking $\mu \equiv m - (1 + a g + b g )$ and $q=q^*$, the early study \cite{Silva:2006xv} of thermodynamics above the BPS limit involves temperature alone.

We previously determined that the potentials are constants $\Phi=\Phi^*=3$ and $\Omega_{1,2}=\Omega^*_{1,2}=g$ on the BPS surface. The combination of potentials,
\be
\label{eqn:varphidef}
  \varphi = \Phi - \frac{\Omega_1 + \Omega_2}{g} - 1~,
\ee
therefore vanishes there and otherwise gives a physical measure of the ``distance'' away from the BPS configurations. Departure of the potentials from 
their BPS values due to a small {\it temperature} cancels from the expression due to \eqref{eqn:potTdiff} so the variable $\varphi$ measures distance {\it along} the extremal surface, independently of temperature. 
The general expressions for $\Phi$, $\Omega_{1, 2}$ \eqref{eqn: potential} give:
\be
  \varphi = \frac{(3 r_+^2 - r^{*2}) (q - q^*) - (1 + a g + b g) (r_+^2 - r^{*2})^2}{(r_+^2 + a^2) (r_+^2 + b^2) + a b q}\, ,
\ee
after some rewriting. The second term in the numerator is negligible near the BPS surface where \eqref{eq:ExtNearBPSVar} instructs us to take $q - q^*$ and $r_+^2 - r^{*2}$ small and of the same order. We find 
\be
\label{eqn:varphiqq}
  \varphi = \frac{2 (q - q^*)}{q^*}\, ,
\ee
at leading order. Therefore the difference $q - q^*$ is a good measure of departures from the BPS surface that preserve extremality. 

To summarize so far, within the extremal surface we can take
\be
\label{eqn:mgrela}
  m = (1 + a g + b g) q~,
\ee
at the linear order and we can relate the horizon coordinate $r^2_+-r^2_*$ to $q-q^*$ through the condition that the temperature \eqref{eqn:Tapx} vanishes. This leaves $q-q^*$ as the only variable that is sensitive to the deviation from the BPS surface. It is equivalent to $\varphi$ through \eqref{eqn:varphiqq}.
The remaining variables $(a, b)$ parametrize the base point on the BPS surface, equivalent to $Q$, $J_{1,2}$ subject to the constraint \eqref{eqn: Charge Constraint}.

After this lengthy discussion of principles, we can return to the general formulae \eqref{eqn: potential} for the potentials $\Phi$, $\Omega_{1, 2}$. In each expression we take fixed $(a,b)$ and expand around $r^2=r^2_*$, $q=q^*$. We then eliminate $r^2_+-r^2_*$ in favor of $q-q^*$ by imposing vanishing temperature and introduce $\varphi$ through \eqref{eqn:varphiqq}. The resulting deviations away from the BPS values
$\Phi^*=3$ and $\Omega^*_{1,2}=g$ can be presented at linear order in $\varphi$ as derivatives of the potentials with respect to $\varphi$:
\begin{align}
  \frac{\partial \Phi}{\partial \varphi} & =  {3(a+b)(2 + (a+b)g) \over 2 [ 1 + 3(a+b)g + (a^2 + 3ab + b^2)g^2]} \\
  & = 1 -  \frac{1}{4} N^2 \frac{J_1 + J_2 + 2 (3 Q \ell_5 + \frac{1}{2} N^2)}{\left(\frac{S^*}{2 \pi} \right)^2 + 2 (3 Q \ell_5 + \frac{1}{2} N^2)^2}\, ,\label{eq:dPhidvarphi}
  \end{align}
and
\begin{align}
\frac{\partial \Omega_1\ell_5}{\partial \varphi} & = -  { (1-a^2 g^2)g \over 2 [ 1 + 3(a+b)g + (a^2 + 3ab + b^2)g^2]} \\
& = - \frac{1}{4} N^2\frac{J_2 + 3 Q \ell_5 + \frac{1}{2} N^2 }{\left(\frac{S^*}{2 \pi} \right)^2 + 2 (3 Q \ell_5 + \frac{1}{2} N^2)^2}\, .\label{eq:dOmegaadvarphi}
\end{align}
The analogous expression for $\partial_\varphi\Omega_2$ is obtained by the substitutions $a \leftrightarrow b$ and $J_1\leftrightarrow J_2$. As a consistency check, we have
\begin{equation}
\label{eqn:varphider}
\partial_\varphi\Phi-\partial_\varphi(\Omega_1+\Omega_2)\ell_5 =1
\end{equation}
as expected from the definition \eqref{eqn:varphidef}.
Having computed the potentials $\Phi - \Phi^*$ and $\Omega_{1,2} - \Omega_{1,2}^*$ at linear order in $\varphi$, we turn to the first law of thermodynamics
\be\label{eq:ExtNearBPSdS}
0 =    T d S = d(M - M^*) - (\Phi - \Phi^*) d Q - (\Omega_1 - \Omega_1^*) d J_1 - (\Omega_2 - \Omega_2^*) d J_2\, ,
\ee
where $T=0$ because we consider extremal black holes. The charges are given by the general expressions (\ref{eqn: charge}-\ref{eqn: angular momenta}) with $m$ eliminated in favor of $q$ through \eqref{eqn:mgrela}. The differentials $dQ$, $dJ_{1,2}$ therefore become linear combinations of $dq$ and $da, db$. We are only interested in the first of these, because the others correspond to motion within the BPS surface. Introducing $\varphi$ though 
\eqref{eqn:varphiqq} we then find:
\be\label{eq:CTCE}
  - (\Phi - \Phi^*) dQ - (\Omega_1 - \Omega_1^*) dJ_1 - (\Omega_2 - \Omega_2^*) dJ_2 = -    {C_\varphi\over T}\frac{\varphi \, d\varphi}{(2 \pi\ell_5)^2} ~,
  \ee
with the temperature independent combination ${C_\varphi\over T}$ given by
\begin{align}
{C_\varphi\over T}   & = \frac{\pi}{4 G_5} \frac{\pi^2 (a+b)^2}{g^2 (1 - a g) (1 - b g)} \frac{3 + (a + b) g - a b g^2}{1 + 3 (a + b) g + (a^2 + b^2 + 3 a b) g^2}\, .\label{eq:CT/T}
\end{align}

After integration of the first law \eqref{eq:ExtNearBPSdS} we have
\be\label{eq:ExtNearBPSM-M*}
        M-M^* = \frac{1}{2}{C_\varphi\over T} \left( \frac{\varphi}{2\pi\ell_5}\right)^2~,
\ee
to the leading order.
The identical result can be derived directly from the general mass formula \eqref{eqn:Mmrelation}. Indeed, the central manipulation is given already in \eqref{eq:horizon recast}, with the 2nd term in the numerator absent when the temperature vanishes $T=0$.

We refer to $C_\varphi$ as the {\it capacitance}. This is the terminology used in standard electromagnetism insofar as $\varphi$ can be identified with the electric potential. The capacitance quantifies the energy 
\eqref{eq:ExtNearBPSM-M*} required to violate the constraint by an amount measured by the potential $\varphi$. Physically, this is quite distinct from the heat capacity $C_T$, a measure of the energy needed to increase the temperature. It is therefore surprising that
$$
C_\varphi = C_T~,
$$
for the black holes we consider. 

We have repeatedly invoked the intuition that the potential $\varphi$ introduced in \eqref{eqn:varphiqq} measures the violation of the constraint \eqref{eqn: Charge Constraint} that must be 
satisfied by all BPS configurations. To make this precise we define the height function:
\be
\label{eqn:hdeff}
  h \equiv (Q\ell_5)^{3} + {1\over 2}N^2 J_1 J_2 - \left(3Q\ell_5 + {1\over 2}N^2 \right)\left(3(Q\ell_5)^2- {1\over 2} N^2 (J_1 + J_2)\right)\, ,
\ee
that quantifies the distance from the constraint surface $h=0$ explicitly. The differential form 
\begin{equation}
\label{eqn:dhdel}
dh = 3 \left( - 8(Q\ell_5) ^2 - N^2 Q\ell_5  + {1\over 2} N^2 (J_1 + J_2)\right) dQ \ell_5 +{1\over 2} N^2  \left( J_2 + 3Q\ell_5 + {1\over 2} N^2  \right) dJ_1 + J_1 \leftrightarrow J_2~,
\end{equation}
realizes the surfaces $h = \textrm{const}$ as 2D planes that are generated by the three one-forms $dQ, dJ_1, dJ_2$ subject to the constraint $dh=0$. In this construction $h$ literally measures the distance along the normal to the 2D constraint surface $h=0$. The BPS surface can be viewed as the intersection of the constraint surface $h=0$ with the extremal surface $T=0$.

Near the BPS surface we can eliminate the parameter $m$ through \eqref{eqn:mgrela} and then (\ref{eqn: charge}-\ref{eqn: angular momenta}) express
the three charges $Q$, $J_{1,2}$ as functions of the parameters $(q, a, b)$. On general grounds, the differential form $dh$ above becomes a linear combination of $dq, da, db$ 
after inserting these formulae for the charges. Explicit computation gives
$$
dh =  \left[8 (Q\ell_5)^3 + \frac{1}{4} N^4 (J_1 + J_2) \right]  {dq\over q^*}~.
$$
The absence of $da, db$ in this formula shows that the curve $dh$ has no components along the BPS surface, as expected. The nontrivial component along $dq$ relates the normalization of $h$ and $\varphi$ through \eqref{eqn:varphiqq}. We find
\be
\label{eqn:hvsvarphi}
  h = \frac{1}{2} \left[8 (Q\ell_5)^3 + \frac{1}{4} N^4 (J_1 + J_2) \right] \varphi\, .
\ee

In the next subsection we will uncover a term in the entropy that is proportional to $\varphi$ with a positive coefficient. Stability therefore motivates us to focus on the halfline where both height functions are nonnegative
\be
\varphi\geq 0~. 
\ee
%


\subsection{General NearBPS Limit}\label{sec: near BPS}

In this subsection we consider the general nearBPS regime where deviations from the BPS surface may have neither fixed charge nor vanishing temperature. 
Some aspects of this amount to reconsidering the effects uncovered in the previous two subsections at the same time. However, their interplay gives important new insights. 

The energy of excitations is the thermodynamic quantity that is conceptually most straightforward in the nearBPS limit. A good starting point is the general mass formula \eqref{eqn:Mmrelation}. It depends on the combination of parameters $m-(1+(a+b)g)q$ that was rewritten in \eqref{eq:horizon recast} without invoking any assumptions black hole variables. Inserting the first order expressions \eqref{eqn:Tapx} for the temperature $T$ and \eqref{eqn:varphiqq} for the potential $\varphi$ we immediately find the excitation energy above the BPS bound: 
\be
M-M^* = \frac{1}{2}\frac{C_{T}}{T}\bigg|_{\text{nearExt}}\left[T^2+\left( \frac{\varphi}{2\pi\ell_5}\right)^2\right]~,\label{eqn: delta M minus charges}
\ee
at quadratic order. We have highlighted the identity of the heat capacity and the electric capacitance by avoiding reference to the latter altogether. The new fact uncovered by the general nearBPS limit is the {\it absence} of any cross-terms $T \varphi$ in the mass formula \eqref{eqn: delta M minus charges}. This demonstrates some sort of rotational symmetry that must be present in the regime we explore. The existence of a continuous structure is much stronger than the equality of $C_T$  and $C_\varphi$ that we stressed in the preceding subsection. 

We next consider the additional entropy due to simultaneously allowing small temperature and violation of the constraint. Starting from the general expression \eqref{eqn: entropy} for the entropy, we apply the procedure explained around \eqref{eqn:mgrela}. Thus we first eliminate the parameter $m$ using \eqref{eqn:mgrela} and expand to linear order in $q-q^*$ and $r^2_+-r^2_*$. We then take appropriate linear combinations so those two variables are eliminated in favor of the temperature $T$ \eqref{eqn:Tapx} and the potential $\varphi$ \eqref{eqn:varphiqq}. These steps give an entropy of the excitations taking the form 
\be
\label{eqn:SmSstar}
  S - S^* = \frac{C_T}{T} T + \frac{C_E}{T} \frac{\varphi}{2 \pi}\, , 
\ee
where the heat capacity $C_T$ agrees with the expression \eqref{heat capacity} found previously by considering temperature on its own and 
\begin{eqnarray}
\label{eqn:CETresult}
  {C_E\over T} &=& {2\pi (Q\ell_5)^2\over S^*} \frac{\pi^2}{g}  {(1 + 2(a+b)g + abg^2) (3+(a+b)g - abg^2) \over 1 + 3(a+b)g+g^2(a^2 + b^2 + 3ab)} \cr
  &=& {2\pi\over S^*} \left({C_T\over T}\right) (3Q + {1\over 2}N^2)~  \, .
\end{eqnarray}

The value of $C_E$ is subject to a subtle ambiguity. The expression \eqref{eqn:SmSstar} gives the entropy $S-S^*$ that is {\it in excess} of the BPS entropy $S^*$. However, the BPS entropy is not a proper function of charges, it is only defined modulo the constraint \eqref{eqn: Charge Constraint}. This caveat is inconsequential on the BPS surface where the constraint is satisfied identically. In contrast, the constraint is proportional to $\varphi$ so, for the additional entropy $S-S^*$ the ambiguity can shift the coefficient of $C_E$ arbitrarily, potentially rendering this quantity unphysical.

This issue must be addressed consistently in computations. For example, we variously express the BPS entropy $S^*$ as a function of parameters $(a,b)$ \eqref{eqn: BPS entropy-1} or as a function of the 
charges $Q$, $J_{1,2}$ \eqref{eqn: BPS entropy}. The differential $dS^*$ computed from the former only gives terms proportional to $da$ and $db$ but when it is evaluated 
from the latter we get terms of the form $dQ$, $dJ_{1,2}$ that, because charges depend on all of the parameters $(q, a, b)$, also yield the differential $dq$. The two forms of the BPS 
entropy therefore give different coefficients in front of the term $d(q-q^*)= 2q^* d\varphi$ even though they agree if we impose the BPS relation $q=q^*$ before computing the differentials. 

We ``gauge fix" the ambiguity by insisting that the BPS entropy takes the canonical form \eqref{eqn: BPS entropy} in terms of $Q$, $J_{1,2}$, rather than something that is equivalent to this formula upon imposing the constraint. Our result for $C_E$ \eqref{eqn:CETresult} is predicated on this convention. 

The final aspects of the general nearBPS limit that we consider are the potential terms in the first law of thermodynamics: 
\be\label{eqn: general ext delta S: 1}
   T dS = d(M - M^*) - (\Phi - \Phi^*) d Q - (\Omega_1 - \Omega_1^*) d J_1 - (\Omega_2 - \Omega_2^*) d J_2\, .
\ee
We can compute the potentials using the procedure described around \eqref{eqn:mgrela}, as in the previous examples. However, for the potentials the results for the general nearBPS limit can equally be inferred from the near extremal ($T\neq 0$ and $\varphi = 0$) and nearBPS extremal ($T = 0$ and $\varphi \neq 0$) special cases that we 
studied in the last two subsections. For example, combining \eqref{eq:dOmegaadT} and \eqref{eq:dOmegaadvarphi} we find the correct result
\begin{eqnarray}
\label{eqn:Omega1expl}
(\Omega_1 - \Omega_1^*)\ell_5  &=& {1\over 4} N^2~\frac{ \left[ J_2 + (3 Q \ell_5+\frac{1}{2} N^2)\right] \varphi  - 
{(2\pi)^2\over S^*} \left[ J_2\left(3 Q \ell_5+\frac{1}{2} N^2\right) -\left(\frac{S^*}{2\pi}\right)^2\right]T } {\left(\frac{S^*}{2 \pi} \right)^2 + \left(3 Q \ell_5 + \frac{1}{2} N^2\right)^2}~,~~~~~~~
\end{eqnarray}
for the angular velocity in the general nearBPS limit. 
The analogous equation for $\Omega_2 - \Omega_2^*$ follows by taking $J_1\leftrightarrow J_2$. The one for $\Phi - \Phi^*$ can be computed similarly from \eqref{eq:PhiadT} and \eqref{eq:dPhidvarphi} 
or by invoking the sum rule
\begin{equation}
\label{eqn:sumrule}
(T \partial_T + \varphi\partial_\varphi ) \left[ \Phi  - (\Omega_1+\Omega_2)\ell_5 \right] = \varphi~,
\end{equation}
that consolidates \eqref{eqn:potTdiff} and \eqref{eqn:varphider} for derivatives with respect to $T$ and $\varphi$, respectively. 

The explicit formula \eqref{eqn:Omega1expl} for $\Omega_1 - \Omega_1^*$ and its analogues for other nearBPS potentials 
are somewhat lengthy and not very illuminating independently. However, they become more instructive when considered together, as a vector in the space of charges. The first law of thermodynamics \eqref{eqn: general ext delta S: 1} motivates construction of the differential form
\begin{eqnarray}
\label{eqn:dhinDirstLaw}
&& T dS^* + (\Phi - \Phi^*) d Q + (\Omega_1 - \Omega_1^*) d J_1 + (\Omega_2 - \Omega_2^*) d J_2 \cr
&=& 
\left( {2\pi\over S^*} (3Q + {1\over 2}N^2) T  +  {\varphi\over 2\pi}     \right) { 2\pi dh \over 
2\left[ \left( {S^*\over 2\pi}\right)^2 + (3Q + {1\over 2}N^2 )^2\right] }~,
\end{eqnarray}
where $dh$ is the one-form \eqref{eqn:dhdel} generated by the height function. Thus the relative strength of the three potentials is precisely the same as the one appearing in the height function $h$, except for the components taken into account by $dS^*$ that are within the BPS surface and only relevant for BPS physics. 

Introducing $\varphi$ using \eqref{eqn:hvsvarphi} the potential terms become
\begin{equation}
\label{eqn:firstlwasfinal}
TdS^* + (\Phi - \Phi^*) d Q + (\Omega_1 - \Omega_1^*) d J_1 + (\Omega_2 - \Omega_2^*) d J_2 = \left[ - \left({C_E\over T}\right) T +  \left({C_T\over T}\right) \varphi\right] d\varphi~,
\end{equation}
where $C_T$ and $C_E$ agree with the functions of charges defined in (\ref{eq:CT/T}) and (\ref{eqn:CETresult}). This is precisely what is needed to satisfy the first law of thermodynamics. Importantly, the values for $C_E$ match only when using the canonical form \eqref{eqn: BPS entropy} of the BPS entropy $S^*$ when evaluating $TdS^*$ in \eqref{eqn:firstlwasfinal}. This illustrates the necessity of treating the ambiguity discussed below \eqref{eqn:CETresult} consistently.

\section{The BPS Partition Function}\label{sec:FreeFT}

In this section, we review the recent progress on the microscopic origin of the AdS$_5$ black hole entropy. We focus on the free field approach applied to $\mathcal{N}=4$ SYM and 
closely follow ~\cite{Kinney:2005ej, Choi:2018hmj}. 

In this and subsequent sections we adopt microscopic units with $\ell_5 = g^{-1} = 1$ and eliminate all references to the Newton's constant $G_5$ 
in favor of the rank of the gauge group $N$ through \eqref{eq:ParamRel}.

\subsection{The Partition Function}
\label{sec:partfct}

Following recent work, we seek to compute the physical partition function
\be\label{eq:physicalZ}
  Z (\beta,\, \Delta_I,\, \omega_i) = \textrm{Tr} \left[e^{- \beta E} e^{\Delta_I Q_I+\omega_i J_i} \right]\, .
\ee
The electric charges and the angular momenta are denoted by $Q_I$ and $J_i$ and the corresponding chemical potentials are $\Delta_I$ and $\omega_i$. Sums over repeated indices $I$ and $i$ are implied. We stress that we do not consider the supersymmetric index since no grading $(-1)^F$ has been inserted. 

The fermionic symmetries $\mathcal{Q}$ and $\mathcal{S}$ transform as spinors under both the $SU(4)_R$ symmetry and the $SO(4)$ little group. We denote the quantum numbers of the supercharge $\mathcal{Q} \equiv \mathcal{Q}_{-t_1, -t_2}^{s_1, s_2, s_3}$ with respect to these groups $\frac{1}{2} s_I$ and $-\frac{1}{2} t_i$ where $s_I=\pm 1$ and $t_i=\pm 1$. The signs of the spinorial indices are explicitly flipped for the conformal supercharges $\mathcal{S} \equiv \mathcal{S}_{t_1, t_2}^{-s_1, -s_2, -s_3}$ which have opposite chirality so we can take the products $\prod s_I \prod_i t_i=+1$ in both cases. 

We write the superalgebra in component form as 
\be
  \{\mathcal{Q},\, \mathcal{S}\} = E - \sum_{I=1}^3 s_I Q_I - \sum_{i=1}^2 t_i J_i\, .
\ee
The anticommutator $\{\mathcal{Q},\, \mathcal{S}\} = 0$ when acting on $\frac{1}{16}$-BPS states so  we have the BPS condition
\be
  E = \sum_{I=1}^3 s_I Q_I + \sum_{i=1}^2 t_i J_i\, .
\ee
We consider the sector where all $s_I, t_i=+1$ without loss of generality. 

The quantum numbers $Q_I, J_i$ are half-integer valued for all states in the theory so the corresponding chemical potentials satisfy the periodicity conditions
\be
 \Delta_I \equiv \Delta_I + 4 \pi i\, ,\quad \omega_i \equiv \omega_i + 4 \pi i\, .
\ee
However, the operator $\mathcal{O} = e^{- \Delta \cdot Q - \omega \cdot J}$ that is inserted in the partition function \eqref{eq:physicalZ} generally does not anticommute with the projection operator $\Gamma$ onto the BPS states
\be
  \mathcal{O} \Gamma = e^{-\frac{1}{2} s\cdot \Delta + \frac{1}{2} t\cdot \omega}\, \Gamma \mathcal{O}\, .
\ee
Anticommutation implies the additional condition:
\be\label{eq:GeneralConstraint}
  s\cdot \Delta  - t\cdot \omega = 2 \pi i\quad (\textrm{mod } 4 \pi i)\, .
\ee
Therefore supersymmetry demands that the potentials satisfy
\be\label{eq:BPSconstr}
  \Delta_1 + \Delta_2 + \Delta_3 - \omega_1 - \omega_2 = 2 \pi i\quad (\textrm{mod } 4 \pi i)\, ,
\ee
for projection to the BPS sector we focus on. 

The complex condition \eqref{eq:BPSconstr} on the potentials is essential. It is closely related to the supersymmetric index, since insertion of $(-)^F$ in the partition function is equivalent 
to the shift $\omega_i\to\omega_i+2\pi i$ for either $i=1$ or $i=2$. However, following recent literature we will maintain reference to the partition function rather than the supersymmetric index and we will consider complex potentials. 
In this terminology the partition function counts protected states (is independent of $\beta$) on the surface defined by the complex supersymmetry condition \eqref{eq:BPSconstr}.


\subsection{Single Particle Enumeration}

We now turn to the central problem of computing the partition function \eqref{eq:physicalZ} of $\mathcal{N}=4$ SYM on $S^1 \times S^3$. We impose 
anti-periodic boundary conditions for fermions along $S^1$, as usual for any partition function but in contradistinction to the supersymmetric index. 
We work perturbatively at weak coupling and express the result as a matrix 
model following \cite{Sundborg:1999ue,Aharony:2003sx, Kinney:2005ej}.

The first step is to enumerate ``single-particle states", in the terminology of the bulk AdS$_5$ theory. In the quantum field theory description these are the individual operators that generate the operator algebra. They can be realized as elementary fields, possibly with derivatives since those do not change the particle number. It is useful to
decompose the field content of $\mathcal{N}=4$ SYM under an $\mathcal{N}=1$ subalgebra and represent is matter as
an $\mathcal{N}=1$ vector multiplet, three $\mathcal{N}=1$ chiral multiplets, and three $\mathcal{N}=1$ anti-chiral multiplets. 
In components, a $\mathcal{N}=1$ vector multiplet contains a gauge boson and a real spinor, while a $\mathcal{N}=1$ chiral multiplet contains a complex scalar and a real chiral spinor. 

We first consider a chiral multiplet with the spectrum:
\vspace{2mm}
\begin{center}
\begin{tabular}{c|c}
  Fields & $(E,\, J_1,\, J_2;\, Q_1,\, Q_2,\, Q_3)$\\
  \hline
  $X = \frac{1}{\sqrt{2}} (\phi^1 + i \phi^2)$ & $(1,\, 0,\, 0;\, 1,\, 0,\, 0)$\\
  $Y = \frac{1}{\sqrt{2}} (\phi^3 + i \phi^4)$ & $(1,\, 0,\, 0;\, 0,\, 1,\, 0)$\\
  $Z = \frac{1}{\sqrt{2}} (\phi^5 + i \phi^6)$ & $(1,\, 0,\, 0;\, 0,\, 0,\, 1)$\\
  \hline
  $\bar{\psi}_{\dot{\alpha},\, a}$ & $(\frac{3}{2},\, 0,\,  \pm \frac{1}{2};\, - \frac{1}{2},\, + \frac{1}{2},\, + \frac{1}{2})$\\
  {} & $(\frac{3}{2},\, 0,\,  \pm \frac{1}{2};\, + \frac{1}{2},\, - \frac{1}{2},\, + \frac{1}{2})$\\
  {} & $(\frac{3}{2},\, 0,\,  \pm \frac{1}{2};\, + \frac{1}{2},\,  + \frac{1}{2},\, - \frac{1}{2})$\\
  \hline
\end{tabular}
\end{center}
The corresponding partition functions become
\begin{align} \label{eqn:chiralsingleb}
  f_B^c (\beta,\, \Delta_I,\, \omega_i) & = \sum_{I=1}^3 e^{\Delta_I}\frac{e^{-\beta} (1 - e^{-2 \beta})}{(1 - e^{-\beta+\omega_1}) (1 - e^{-\beta+\omega_2}) (1 - e^{-\beta-\omega_1}) (1 - e^{-\beta-\omega_2})}\, ,\\
  f_F^c (\beta,\, \Delta_I,\, \omega_i) & = \sum_{I=1}^3 e^{\Delta_I}\frac{e^{-\frac{3}{2} \beta - \Delta} \left((e^{\omega_+} + e^{-\omega_+}) - e^{-\beta} (e^{\omega_-} + e^{-\omega_-}) \right)}{(1 - e^{-\beta+\omega_1}) (1 - e^{-\beta+\omega_2}) (1 - e^{-\beta-\omega_1}) (1 - e^{-\beta-\omega_2})}\, ,
  \label{eqn:chiralsinglef}
\end{align}
where
\be\label{eq:DefineDeltaOmega}
  \Delta \equiv \frac{\Delta_1 + \Delta_2 + \Delta_3}{2}\, ,\quad \omega_\pm \equiv \frac{\omega_1 \pm \omega_2}{2}\, .
\ee
The common denominators of (\ref{eqn:chiralsingleb}-\ref{eqn:chiralsinglef}) incorporate the quantum numbers of the four components in the gradient $\nabla_\mu$, each resummed as a geometric series to take into account any number of derivatives. The positive terms in the numerators encode the data from the table, while the negative ones correspond to subtraction of operators that satisfy their equations of motion.

The anti-chiral multiplet can be obtained from the chiral multiplet by
\be
  J_1 \leftrightarrow J_2\, ,\quad Q_I \to - Q_I\, .
\ee
For bosons this just changes the overall factor giving the R-charge, but for fermions the Lorentz indices of the fields and the equation of motion exchange their 
chirality $\omega_+\leftrightarrow\omega_-$, in addition to the R-charges changing sign:
\begin{align}
  f_B^a (\beta,\, \Delta_I,\, \omega_i) & = \sum_{I=1}^3 e^{-\Delta_I}\frac{e^{-\beta} (1 - e^{-2 \beta})}{(1 - e^{-\beta+\omega_1}) (1 - e^{-\beta+\omega_2}) (1 - e^{-\beta-\omega_1}) (1 - e^{-\beta-\omega_2})}\, ,\label{eq:fBa} \\ 
  f_F^a (\beta,\, \Delta_I,\, \omega_i) & = \sum_{I=1}^3 e^{- \Delta_I}\frac{e^{-\frac{3}{2} \beta + \Delta} \left((e^{\omega_-} + e^{-\omega_-}) - e^{-\beta} (e^{\omega_+} + e^{-\omega_+}) \right)}{(1 - e^{-\beta+\omega_1}) (1 - e^{-\beta+\omega_2}) (1 - e^{-\beta-\omega_1}) (1 - e^{-\beta-\omega_2})}\, .\label{eq:fFa}
\end{align}

The vector multiplet has the spectrum:
\vspace{2mm}
\begin{center}
\begin{tabular}{c|c}
  Fields & $(E,\, J_1,\, J_2;\, Q_1,\, Q_2,\, Q_3)$\\
  \hline
  $A_\mu\, (\mu = 1,\,\cdots,\, 4)$ & $(1,\, \pm 1,\, \pm 1;\, 0,\, 0,\, 0)$\\
  \hline
  $\psi^a_\alpha$ & $(\frac{3}{2},\, \pm \frac{1}{2},\, 0;\, + \frac{1}{2},\, + \frac{1}{2},\, + \frac{1}{2})$\\
  \hline
  $\bar{\psi}_{\dot{\alpha},\, a}$ & $(\frac{3}{2},\, 0,\,  \pm \frac{1}{2};\, - \frac{1}{2},\, - \frac{1}{2},\, - \frac{1}{2})$\\
  \hline
\end{tabular}
\end{center}
with the single particle partition functions:
\begin{align}
  f_B^v (\beta,\, \Delta_I,\, \omega_i) & = \frac{e^{-\beta}  (e^{\omega_1} + e^{\omega_2} + e^{-\omega_1} + e^{-\omega_2}) - 1 - e^{-2 \beta}}{(1 - e^{-\beta+\omega_1}) (1 - e^{-\beta+\omega_2}) (1 - e^{-\beta-\omega_1}) (1 - e^{-\beta-\omega_2})} (1 - e^{-2 \beta}) + 1\, ,\label{eq:fBv}\\
  f_F^v (\beta,\, \Delta_I,\, \omega_i) & = \frac{e^{-\frac{3}{2} \beta} (e^{\Delta} - e^{-\Delta} e^{-\beta}) (e^{\omega_+} + e^{-\omega_+}) + e^{-\frac{3}{2} \beta} (e^{-\Delta} - e^{\Delta} e^{-\beta}) (e^{\omega_-} + e^{-\omega_-})}{(1 - e^{-\beta+\omega_1}) (1 - e^{-\beta+\omega_2}) (1 - e^{-\beta-\omega_1}) (1 - e^{-\beta-\omega_2})}\, .\label{eq:fFv}
 \end{align}
The terms in the boson numerator indicate the vector field $A_\mu$ (that has the same quantum numbers as $\nabla_\mu$ in the denominator), with subtractions for the gauge function $\Lambda$ and the Lorentz condition $\nabla_\mu A^\mu=0$. The overall factor $(1-e^{-2\beta})$ imposes the Klein-Gordon equation on all operators and the additional $+1$ corrects for the fact we 
erroneously counted $\Lambda$ with no derivative acting on it as a pure gauge degree of freedom. The gaugini follow from the fermions in chiral and anti-chiral multiplets, upon omission of the overall $\sum_{I=1}^3 e^{-\Delta_I}$ and subsequent reversal of the correlation between chirality and R-charge. 

In order to obtain the BPS condition
\be\label{eq:ExtremCond}
  E = \sum_I Q_I + \sum_i J_i~,
\ee
easily in the low temperature limit $\beta \to \infty$, we rewrite the physical partition function \eqref{eq:physicalZ} 
\begin{align}
\label{eqn:physZstar}
  Z (\beta,\, \Delta_I,\, \omega_i)  & = \textrm{Tr} \left[e^{- \beta (E - \sum_I Q_I - \sum_i J_i)} e^{\widetilde{\Delta}_I  Q_I +\widetilde{\omega}_i J_i} \right]\, ,
\end{align}
where \footnote{It would be proper to use the potentials that have tilde in our notation in nearly all applications of thermodynamics to BPS black holes. We were sloppy on this point when displaying the BPS free energy \eqref{eq:FreeEnergy} in the introduction and in the algebraic manipulations in subsection \ref{sec:partfct}, but not in this subsection prior to this point. We have opted for some ambiguity on the terminology in order to avoid heavy notation where it is not likely to cause confusion.}
\be
\label{eqn:tildevardef}
  \widetilde{\Delta}_I \equiv \Delta_I - \beta\, ,\quad \widetilde{\omega}_i \equiv \omega_i - \beta~.
\ee
Thence only the states satisfying the extremality condition \eqref{eq:ExtremCond} contribute to the partition function in
the limit $\beta\to\infty$ with fixed $\widetilde{\Delta}_I$ and $\widetilde{\omega}_i$. 

The total single particle partition function at $\beta=\infty$ for all fields in ${\cal N}=4$ SYM is exceptionally simple, because only a few single particle operators contribute. 
Considering bosons and fermions separately, we have
\begin{eqnarray}
\label{eqn:fBfinal}
f_B  &=& f_B^v  + f_B^c + f_B^a =   {\sum_I e^{-\tilde{\Delta}_I}  + e^{-\tilde{\omega}_1-\tilde{\omega}_2} \over (1 - e^{- \tilde{\omega}_1})(1 - e^{- \tilde{\omega}_2})}    ~,\cr
f_F & = &  f_F^v  + f_F^c + f_F^a =
{  \sum_I e^{\tilde{\Delta}_I - \tilde{\Delta}}e^{-\tilde{\omega}_+} + e^{-\tilde{\Delta} + \tilde{\omega}_+} \over (1 - e^{- \tilde{\omega}_1})(1 - e^{- \tilde{\omega}_2})} - 
e^{-\tilde{\Delta}+\tilde{\omega}_+}  ~.
\label{eqn:fFfinal}
\end{eqnarray}
The general expressions for any $\beta$ are much more complicated. However, the single particle partition function with the constraint $e^{-\tilde{\Delta} + \tilde{\omega}_+}=-1$ imposed
\begin{eqnarray}
\label{eqn:singeletter}
f_B + f_F &=&  1 - { \prod_I ( 1  - e^{-\tilde{\Delta}_I})  \over (1 - e^{- \tilde{\omega}_1} )(1 - e^{- \tilde{\omega}_2})} ~,
\end{eqnarray}
is in fact {\it independent} of $\beta$. This independence follows from the general arguments familiar from the Witten index and can also be verified by explicit computation. Equivalently, the supersymmetric index $f_B-f_F$ is temperature 
independent on the constraint surface $e^{-\tilde{\Delta} + \tilde{\omega}_+}=+1$. In fact, it is identical to \eqref{eqn:singeletter}. Therefore, for these quantities the simple result found 
at low temperature $\beta=\infty$ applies at any temperature.

\subsection{Multiparticle Enumeration}

All of the ``letters" realized by single particle operators can be multiplied together to form ``words". Starting from any of the operators enumerated by the single particle partition 
function $f(\beta,\, \Delta_I,\, \omega_i)$ we can form $n$-particle products counted by $f(n\beta,\, n\Delta_I,\, n\omega_i)$. An overall trace must be imposed to ensure gauge invariance 
but then $n$-fold cyclicity of the trace must be taken account so, after summing over any number of operators, the single trace partition function becomes
\be
  Z_{\rm ST} = \sum_{k=1}^\infty {f(n\beta,\, n\Delta_I,\, n\omega_i)\over n}  \, .
\ee
Furthermore, having determined all single trace operators, taking multi-trace operators into account exponentiates the counting. The full partition function becomes
\be
\label{eqn:totalZ}
  Z_{\rm MP} (\beta,\, \Delta_I,\, \omega_i)  = \textrm{exp} \left[\sum_{n=1}^\infty \frac{1}{n} \left[f_B (n\beta,\, n\Delta_I,\, n\omega_i) + (-1)^{n+1}\, f_F  (n\beta,\, n\Delta_I,\, n\omega_i)\right]  \right]\, .
\ee
This result is purely combinatorial so the statistics of fermions only enter through the exclusion principle. This accounts for the prefactor $(-1)^{n+1}$ in front of the fermionic terms.

The brief justification of the multiparticle partition function \eqref{eqn:totalZ} presented in this subsection has been cavalier about the combinatorics. The relative ordering of operators within traces is important for the detailed enumeration and somewhat elaborate combinatorics (Polya theory) must be invoked. However, for large $N$ the correct result is in fact \eqref{eqn:totalZ} and the heuristic arguments give in this subsection serve to motivate the key features of the formula. 

\subsection{The $\mathcal{N}=4$ SYM Perturbative Matrix Model}
One additional feature must taken into account. All of the quantum fields transform in the adjoint of $U(N)$ and we must incorporate the gauge indices. 
We (somewhat prematurely) imposed a singlet condition on the operators already in the preceding subsection. Incorporation of the full gauge structure gives the unitary matrix model
\cite{Sundborg:1999ue, Aharony:2003sx, Kinney:2005ej}
\begin{align}
  Z (\beta,\, \Delta_I,\, \omega_i) & = \int [dU]\, \textrm{exp} \Bigg\{\sum_{n=1}^\infty \frac{1}{n} f_n(n\beta,\, n\Delta_I,\, n\omega_i)  \, \chi_{\rm Adj} (U^n)  \Bigg\}\, ,
\end{align}
where $U$ denotes a $U(N)$ matrix, $\chi_{\rm Adj}$ the character in the adjoint representation, and 
\begin{equation}
\label{eqn:fndef}
f_n(n\beta,\, n\Delta_I,\, n\omega_i) = f_B(n\beta,\, n\Delta_I,\, n\omega_i)  + (-1)^{n+1}\, f_F  (n\beta,\, n\Delta_I,\, n\omega_i)~.
\end{equation}
For completeness, we recall that in the case of
weakly coupled $\mathcal{N}=4$ SYM the single particle partition functions are
\begin{eqnarray}
\label{eqn:fBgen}
f_B  &=& f_B^v  + f_B^c + f_B^a   ~,\cr
f_F & = &  f_F^v  + f_F^c + f_F^a  ~,
\label{eqn:fFgen}
\end{eqnarray}
where the constituent functions were given in (\ref{eqn:chiralsingleb}, \ref{eq:fBa}, \ref{eq:fBv}) and (\ref{eqn:chiralsinglef}, \ref{eq:fFa}, \ref{eq:fFv}), respectively.
Their explicit expressions are not illuminating in general but for $\beta=\infty$ they greatly simplify and reduce to \eqref{eqn:fFfinal}.

The standard strategy for integrating over all unitary matrices is to represent them in terms of their eigenvalues $e^{i{\alpha}_{a}}$ 
(so those of the adjoint representation become $e^{i{\alpha}_{ab}}$ where $\alpha_{ab} \equiv \alpha_a - \alpha_b$) 
and change variables to an integral over eigenvalues 
\begin{align}
  Z (\beta,\, \Delta_I,\, \omega_i) & = \frac{1}{N!} \oint \prod_{a=1}^{N} \frac{d \alpha_a}{2 \pi}\, \prod_{a<b} \left(2\, \textrm{sin} \frac{\alpha_{ab}}{2} \right)^2 \, 
  \textrm{exp} \bigg[ \sum_{a, b = 1}^N \sum_{n=1}^\infty \frac{1}{n} f_n(n \beta,\, n\Delta_I,\, n \omega_i)\, e^{i n \alpha_{ab}} \bigg]\, .\label{eq:MatrixModel}
\end{align}
The factor involving $\sin\frac{\alpha_{ab}}{2}$ is the van der Monde determinant that arises as a Jacobian due to the change of variables. 

We stress that the BPS constraint \eqref{eq:BPSconstr} has not yet been imposed on the matrix model so the chemical potentials $\Delta_I$ and $\omega_i$ are still independent parameters
and the partition function \eqref{eq:MatrixModel} includes contributions from nonBPS states.

\subsection{The BPS Limit}

The analysis of matrix models is a highly developed science \cite{Ginsparg:1993is,Marino:2011nm}. The established intuition is that the eigenvalues of the matrices experience a universal repulsion because of the van der Monde determinant that may be balanced by attraction due to a model-dependent potential, which in the current context is closely related to the single particle distribution $f$. The repulsion favors a uniform eigenvalue distribution that corresponds to a confined phase with free energy of order ${\cal O}(1)$ while attraction can prompt localization that gives rise to a deconfined phase where free energy increases to order ${\cal O}(N^2)$ \cite{Gross:1980he}. 

Early studies of the matrix model for ${\cal N}=4$ SYM failed to identify a deconfined phase appropriate for the description of macroscopic black holes \cite{Kinney:2005ej}, but recent 
research make claims to the contrary \cite{Choi:2018vbz}. We will refrain from a nuanced discussion of the evidence one way or another and simply assume a deconfined phase where the phases $e^{i n \alpha_{ab}}$ do not give cancellations at the leading order. Then the matrix model \eqref{eq:MatrixModel} yields the partition function 
$$
\ln   Z (\beta,\, \Delta_I,\, \omega_i) = N^2 \sum_{n=1}^\infty \frac{1}{n} f_n(n \beta,\, n\Delta_I,\, n \omega_i)~.
$$

The BPS limit requires vanishing temperature and we must also impose the BPS constraint \eqref{eq:BPSconstr}. We already computed the single particle partition function $f$ in this 
limit with the result \eqref{eqn:singeletter}. The fermion phase factor $(-)^{n+1}$ in \eqref{eqn:fndef} conspires with signs from the constraint $e^{-\tilde{\Delta} + \tilde{\omega}_+}=-1$ such 
that the multiparticle generalization becomes 
\begin{eqnarray}
\label{eqn:singletter}
f_n(n\beta,\, n\Delta_I,\, n\omega_i)   &=&  1 - { \prod_I ( 1  - e^{-n\tilde{\Delta}_I})  \over (1 - e^{- n\tilde{\omega}_1} )(1 - e^{- n\tilde{\omega}_2})} ~.
\end{eqnarray}
This is exactly the result reported in ~\cite{Choi:2018hmj}. There the analysis focussed on the high temperature limit $\beta \to 0$ while we have discussed the low temperature regime
$\beta \to \infty$. The agreement of the results is due to the temperature independence along the BPS surface. 

We further restrict the discussion to the Cardy limit $|\widetilde{\omega}_i| \ll 1$. In this situation it has been argued that all significant contributions to the sum over $n$ 
are from sufficiently small $n$ that $n |\widetilde{\omega}_i|\ll 1$.  
Therefore
\begin{eqnarray}
\ln   Z (\beta,\, \Delta_I,\, \omega_i) &=& \frac{N^2}{\widetilde{\omega}_1 \widetilde{\omega}_2}  \sum_{n=1}^\infty \frac{1}{n^3}  \prod_I ( 1  - e^{-n\tilde{\Delta}_I})  \cr
& = & \frac{N^2}{\widetilde{\omega}_1 \widetilde{\omega}_2} 
\sum_{s_1 s_2 s_3 = + 1} \left[\textrm{Li}_3 \left(- e^{\frac{s_I \widetilde{\Delta}_I}{2}} \right) - \textrm{Li}_3 \left(- e^{- \frac{s_I \widetilde{\Delta}_I}{2}} \right)\right] ~.
\end{eqnarray}
The identity
\be
  \textrm{Li}_3 (- e^x) - \textrm{Li}_3 (- e^{-x}) = - \frac{x^3}{6} - \frac{\pi^2 x}{6}\quad \textrm{ for } - \pi < \textrm{Im} (x) < \pi\, ,
\ee
finally gives the free energy
\begin{eqnarray} 
\label{eq:KimResult-1}
\ln   Z (\beta,\, \Delta_I,\, \omega_i)  &=& - \frac{N^2}{6\, \widetilde{\omega}_1 \widetilde{\omega}_2} 
\sum_{s_1 s_2 s_3 = +1} \left[ {1\over 8} \left(s_I \widetilde{\Delta}_I\right)^3 + {1\over 2}\pi^2 \left(s_I \widetilde{\Delta}_I \right) \right]\cr
& = & - \frac{1}{2}N^2 \frac{\widetilde{\Delta}_1 \widetilde{\Delta}_2 \widetilde{\Delta}_3}{\widetilde{\omega}_1 \widetilde{\omega}_2}\, .
\end{eqnarray}


\subsection{Discussion}

The partition function \eqref{eq:KimResult-1} with the complex constraint on potentials \eqref{eq:BPSconstr} undoubtedly describes $\mathcal{N}=4$ SYM in the regime relevant for comparison with AdS$_5$ BPS black holes. It was first inferred from black hole thermodynamics \cite{Hosseini:2017mds} and recently derived from the Euclidean path integral and supersymmetric localization \cite{Cabo-Bizet:2018ehj}, from free field analysis \cite{Choi:2018hmj}, and from computation of the superconformal index using Bethe vacua \cite{Benini:2018ywd}. All these works pursue very similar ideas but some aspects of the computations and their interrelation remains unclear, 
at least to the authors of this article.

The simple computations we presented in this section following \cite{Choi:2018hmj} give an interesting free field representation that appears to capture some aspects of physics in the strongly coupled region. The situation is akin to a model that represents a CFT$_2$ with central charge $c$ as a free model such as $c$ free bosons. Such a
toy model of CFT$_2$ generally misses many detailed features of the theory, but it captures some aspects of the CFT$_2$ robustly (such as the Casimir energy and the entropy) and so it serves as a useful benchmark.

In other words, we are not committed to the free field derivation, but the upshot one way or another is that the partition function for the BPS limit is 
\be
\label{eqn:BPSparti}
  \ln~ Z = -\frac{1}{2}N^2 \frac{\widetilde{\Delta}_1 \widetilde{\Delta}_2 \widetilde{\Delta}_3}{\widetilde{\omega}_1 \widetilde{\omega}_2}\, .
\ee
Our goal in the subsequent section is to leverage this result, whichever way it came about, to account also for nonBPS physics.

\section{Black Hole Statistical Physics}\label{sec:BHStatMech}

In this section we discuss thermodynamics of black holes starting from the BPS result for the microscopic free energy \eqref{eqn:BPSparti}. Our emphasis is on thermodynamic variables for nearBPS black holes. 
 
\subsection{Studying nearBPS using BPS data: Introduction}\label{sec:firstlawmicro}

The microscopic considerations in section \ref{sec:FreeFT} studied the partition function \eqref{eqn:physZstar} which we reproduce here for convenience: 
\begin{align}
\label{eqn:physZtar}
  Z (\beta,\, \Delta_I,\, \omega_i)  & = \textrm{Tr} \left[ e^{- \beta (E - E^*)}e^{\widetilde{\Delta}_I  Q_I + \widetilde{\omega}_i J_i} \right]\, . 
\end{align}
The notation strongly suggests that this partition function depends on all the variables $\beta, \Delta_I, \omega_i$ but in fact the computations in section \ref{sec:FreeFT} 
were restricted to the BPS limit, a surface of real codimension two. The restriction on the domain of $Z$ can be specified by the constraints
$\beta=\infty$ and $\sum_I\widetilde{\Delta}_I - \sum_i \widetilde{\omega}_i = 2\pi i$. 

The goal of this section is to generalize the microscopic description beyond BPS and so allow deformations of both these constraints by small amounts. Our strategy is to make mild smoothness assumptions on the partition function, which are then validated by comparison with our results from gravity. To clarify how it is possible to learn anything interesting from such minimal assumptions, it is instructive to consider a simple example.  

As a start, we must address an issue of conventions. In our gravitational computations, the thermodynamic variables satisfy the first law \eqref{eqn: general ext delta S: 1}. 
Comparison with the microscopic partition function \eqref{eqn:physZstar} indicates a relative factor of $\beta$ in the potentials, leading to 
the identifications $\beta(\Phi_I-\Phi^*_I)=\widetilde{\Delta}_I$ and $\beta(\Omega_i-\Omega^*_i)=\widetilde{\omega}_i$.\footnote{The gravitational computations in section \ref{sec:BHThermo} were restricted to the diagonal case where the three potentials are equal but this limitation does not apply to formulae in this section. When we ``compare" with section \ref{sec:BHThermo} we only literally compare for diagonal charges. The generic formula we derive in this section constitute microscopic {\it predictions} for the gravitational side.} The shift of the gravitational potentials by their BPS values $\Phi^*_I$, $\Omega^*_i$ corresponds precisely to the definition \eqref{eqn:tildevardef} of microscopic potentials with a tilde from those without tilde. 

The meticulous tracking of conventions gives an immediate payoff in the extremal limit $T\to 0$. The limit is taken such that variables with tilde are kept fixed. Therefore, the potentials that appear in the microscopic partition function are identified with the {\it thermal derivative} of the potentials employed in the gravity 
description \cite{Silva:2006xv,Kim:2006he}:
\begin{eqnarray}
\label{eqn:reDelderT}
{\rm Re} ~\tilde{\Delta}_I & = & \partial_T\Phi_I~,\cr
{\rm Re} ~\tilde{\omega}_i & = & \partial_T\Omega_i~.
\end{eqnarray}
The microscopic partition function \eqref{eqn:physZtar} gives values for the left hand side that should coincide with the gravitational results on the right hand side. We will verify this expectation in subsection \ref{sec:ReIm}.

The identifications \eqref{eqn:reDelderT} also illustrate the strategy for going beyond BPS. These equations were found by taking the extremal limit $T\to 0$ but supersymmetry was not invoked. This provenance suggests their validity also when the constraint is violated. We will confirm this expectation below. This successful comparison is a simple example of leveraging BPS results to study the nearBPS regime. 

Recall that the supersymmetry condition \eqref{eq:BPSconstr} is complex  and by continuity potentials remain complex in the entire nearBPS region. In our identifications \eqref{eqn:reDelderT} we identified the {\it real} part of the potentials with the corresponding physical field in spacetime. This is justified because the physical charges are real and so their conjugate potentials are real, according to the partition function \eqref{eqn:physZtar}. We will later find more data about the nearBPS by exploiting both the real and imaginary parts of the potential, as well their interplay. 

BPS configurations have energy $E=E^*$ and so the saddlepoint approximation to the partition function \eqref{eqn:physZtar} gives the entropy
\be\label{eq:EntropyFct} 
S = \ln Z  -    \widetilde{\Delta}_I  Q_I - \widetilde{\omega}_i J_i-  \Lambda \Big( \sum_I \widetilde{\Delta}_I - \sum_i \widetilde{\omega}_i - 2 \pi i \Big)~.
\ee
The constraint on potentials \eqref{eq:BPSconstr} was imposed by introducing a Lagrange multiplier $\Lambda$.
The expression \eqref{eq:EntropyFct} is referred to as the {\it entropy function}. Unlike the usual entropy it is a function of potentials but, after they are extremized over, it gives the black hole entropy as function of charges. In the present context there is little more to the terminology than a basic change of thermodynamic ensemble through Legendre transform (but there are impressive generalizations \cite{Sen:2007qy}). In the next subsection we will extremize the entropy function \eqref{eq:EntropyFct} for BPS black holes explicitly. Later, in subsection \ref{sec:nearBPSext}, we will generalize the entire entropy function to the nearBPS regime using minimal assumptions.


\subsection{Entropy Extremization for BPS Black Holes}\label{sec:Extremization}

In this subsection we review the computation of BPS black hole entropy. We start from the entropy function $S$ \eqref{eq:EntropyFct} derived from the BPS partition function \eqref{eqn:BPSparti}. Our analysis mainly follow~\cite{Cabo-Bizet:2018ehj}.

Extremization of the entropy function \eqref{eq:EntropyFct} with $\ln Z$ given by \eqref{eqn:BPSparti} gives 
\begin{align}
\frac{\partial S}{\partial \Delta_I} & = - \frac{1}{2} N^2\frac{1}{\widetilde{\Delta}_I}~\frac{\widetilde{\Delta}_1 \widetilde{\Delta}_2 \widetilde{\Delta}_3}{\widetilde{\omega}_1\widetilde{\omega}_2} - (Q_I + \Lambda) = 0\, ,\label{eq:BPSExtEq-1}\\
\frac{\partial S}{\partial \omega_i} & = \frac{1}{2} N^2\frac{1}{\widetilde{\omega}_i} \frac{\widetilde{\Delta}_1 \widetilde{\Delta}_2 \widetilde{\Delta}_3}{\widetilde{\omega}_1\widetilde{\omega}_2}   - (J_i - \Lambda) = 0\, ,\label{eq:BPSExtEq-2}\\
\frac{\partial S}{\partial \Lambda} & = \sum_I \widetilde{\Delta}_I - \sum_i \widetilde{\omega}_i - 2 \pi i = 0\, .\label{eq:BPSExtEq-3}
\end{align}
The last equation imposes the constraint \eqref{eq:BPSconstr} on $\widetilde{\Delta}_I$ and $\widetilde{\omega}_i$. 

We obtain the BPS entropy by simplifying the entropy function \eqref{eq:EntropyFct} using the extremization conditions (\ref{eq:BPSExtEq-1}, \ref{eq:BPSExtEq-2}):
\be\label{eq:EntropyLambda}
  S^* = 2 \pi i \Lambda\, .
\ee
It is therefore essential to find the Lagrange multiplier $\Lambda$. To do so, we combine \eqref{eq:BPSExtEq-1} and \eqref{eq:BPSExtEq-2} to find 
\be\label{eq:CubicLambda}
  \prod_I (Q_I + \Lambda) + \frac{1}{2} N^2\prod_i (J_i - \Lambda) = 0\, .
\ee
This is a cubic equation that yields the Lagrange multiplier $\Lambda$, and so the BPS entropy $S^*$, as a function of the black hole charges $Q_I$ and $J_i$. The explicit form of the cubic equation is
\be
  \Lambda^3 + A \Lambda^2 + B \Lambda + C = 0~,
\label{eq:CubicExplicit-1}
\ee
where
\begin{align}
\begin{split}
\label{eqn:ADCdef}
  A & = Q_1 + Q_2 + Q_3 + \frac{1}{2} N^2\, ,\\
  B & = Q_1 Q_2 + Q_2 Q_3 + Q_3 Q_1 - \frac{1}{2} N^2 (J_1 + J_2)\, ,\\
  C & = Q_1 Q_2 Q_3 + \frac{1}{2} N^2 J_1 J_2\, .
\end{split}
\end{align}

Imposing reality of the physical entropy \eqref{eq:EntropyLambda} demands a purely imaginary $\Lambda$. Since the charges $Q_I$ and $J_i$ are all real, the purely imaginary 
roots of \eqref{eq:CubicExplicit-1} appear in pairs and the cubic equation must factorize as 
\be\label{eq:cubicfactorize}
 (\Lambda^2 + B)(\Lambda + A) = 0 ~.
\ee
Thus coefficients in the cubic satisfy $C-AB=0$ or
\be\label{eq:ChargeConstraint}
\Bigg(Q_1 Q_2 Q_3 + \frac{1}{2} N^2 J_1 J_2\Bigg)  - \Bigg(Q_1 + Q_2 + Q_3 + \frac{1}{2} N^2\Bigg)  \Bigg(Q_1 Q_2 + Q_2 Q_3 + Q_3 Q_1 - \frac{1}{2} N^2(J_1 + J_2) \Bigg) = 0\, . 
\ee
This is the constraint on black hole charges \eqref{eqn: Charge Constraint} that must be satisfied on the BPS surface, as a consequence of 
the BPS formula for the mass. 
It generalizes the constraint \eqref{eqn: Charge Constraint} previously found from gravity, with perfect agreement when the three charges are identical. When the constraint is satisfied, the root for $\Lambda$ with negative imaginary part gives the BPS entropy 
\begin{equation}
\label{eqn:Slambda}
S^* = 2\pi i \Lambda = 2\pi \sqrt{Q_1 Q_2 + Q_2 Q_3 + Q_3 Q_1 - \frac{1}{2}N^2 (J_1 + J_2) }~. 
\end{equation}
This formula similarly generalizes the result from gravity \eqref{eqn: BPS entropy} with agreement when the three charges are identical.

We also need the potentials $\widetilde{\Delta}_I$ and $\widetilde{\omega}_i$ at the extremum of the entropy function. We consider $\Lambda = {S^*\over 2\pi i}$ the known function of the charges given through \eqref{eqn:Slambda}. The extremization conditions (\ref{eq:BPSExtEq-1}, \ref{eq:BPSExtEq-2}) give
the ratios
\begin{eqnarray}
\label{eqn:Delomratio}
{\widetilde{\Delta}_I \over \widetilde{\omega}_i } &=&  - {J_i - \Lambda \over  Q_I + \Lambda }~,
\end{eqnarray}
for any $I = 1,\, 2,\, 3$ and $i = 1,\, 2$. Comparing \eqref{eq:BPSExtEq-2} with $i=1$ and $i=2$ we also find the ratio
\be
\label{eqn:omegaratio}
{ \widetilde{\omega}_1\over  \widetilde{\omega}_2 } = {J_2 - \Lambda\over J_1 - \Lambda}~.
\ee
The constraint \eqref{eq:BPSconstr}, i.e. the equation of motion for $\Lambda$ \eqref{eq:BPSExtEq-3}, now gives
\be
\label{eqn:keyom}
 {\widetilde{\omega}_1 + \widetilde{\omega}_2 + 2\pi i\over  \widetilde{\omega}_i } =  {\widetilde{\Delta}_1 + \widetilde{\Delta}_2 + \widetilde{\Delta}_3 \over  \widetilde{\omega}_i } 
 =  -  (J_i -\Lambda )\left( { 1 \over  Q_1 +\Lambda } + { 1 \over  Q_2 +\Lambda }+ { 1 \over  Q_3 +\Lambda }\right)~,
\ee
where we have used the ratios $\tilde{\Delta}_I / \tilde{\omega}_i$ given in \eqref{eqn:Delomratio}. The ratio between the $\tilde{\omega}_i$'s \eqref{eqn:omegaratio} let us reorganize as 
\begin{align}
{  2\pi i\over \widetilde{\omega}_i} & = -  (J_1 -\Lambda )\left(  {1\over J_1 - \Lambda} + {1\over J_2 - \Lambda} + 
{ 1 \over  Q_1 +\Lambda } + { 1 \over  Q_2 +\Lambda }+ { 1 \over  Q_3 +\Lambda }\right)~,\label{eq:InverseOmega}\\
{  2\pi i\over \widetilde{\Delta}_I} & = (Q_I +\Lambda )\left(  {1\over J_1 - \Lambda} + {1\over J_2 - \Lambda} + 
{ 1 \over  Q_1 +\Lambda } + { 1 \over  Q_2 +\Lambda }+ { 1 \over  Q_3 +\Lambda }\right)~.\label{eq:InverseDelta}
 \end{align} 
The second line was found by invoking the ratio \eqref{eqn:Delomratio}. The inverses of these equations give
\begin{align}
  \frac{\widetilde{\omega}_i}{2 \pi i} & = \frac{1}{2} N^2\frac{\prod_k (J_k - \Lambda)}{J_i - \Lambda} \frac{1}{2 \Lambda (\Lambda + Q_1 + Q_2 + Q_3 + \frac{1}{2} N^2)}\, ,\label{eq:SolOmegai}\\
  \frac{\widetilde{\Delta}_I}{2 \pi i} & = \frac{\prod_K (Q_K + \Lambda)}{Q_I + \Lambda} \frac{1}{2 \Lambda (\Lambda + Q_1 + Q_2 + Q_3 + \frac{1}{2} N^2)}\, .\label{eq:SolDeltaI}
\end{align}
We used \eqref{eq:CubicLambda} to simplify the algebra. 
These are the explicit results for the potentials on the BPS surface expressed in terms of charges. We recall that $\Lambda = {S^*\over 2\pi i}$ is the known function of the charges given through \eqref{eqn:Slambda}.

\subsection{NearBPS Microscopics}\label{sec:ReIm}
We now want to leverage the microscopic results derived in the BPS limit to study nearBPS black holes as well.  

The real part of the potentials (\ref{eq:SolOmegai}) are
\be
{\rm Re} ~\widetilde{\omega}_1 = - {\pi^2 N^2 \over S^*}  { J_2 (Q_1 + Q_2 + Q_3 + \frac{1}{2}N^2) - \left(\frac{S^*}{2 \pi} \right)^2 \over \left(\frac{S^*}{2 \pi} \right)^2  + (Q_1 + Q_2 + Q_3 + \frac{1}{2}N^2)^2} \, ,\label{eq:ReOmegai}
\ee
with an analogous expression for ${\rm Re} ~\widetilde{\omega}_2$, and the real part of the potentials \eqref{eq:SolDeltaI} similarly are
\begin{align}
  {} & {\rm Re} ~\widetilde{\Delta}_1 = \, - {2\pi^2\over S^*}{ (Q_2 Q_3 - \left(\frac{S^*}{2 \pi} \right)^2) (Q_1 + Q_2 + Q_3 + \frac{1}{2}N^2) + \left(\frac{S^*}{2 \pi} \right)^2 (Q_2 + Q_3)  \over \left(\frac{S^*}{2 \pi} \right)^2 + (Q_1 + Q_2 + Q_3 + \frac{1}{2}N^2)^2 }~, \label{eq:ReDelta1}
\end{align}
with analogous equations for $\textrm{Re}\, \widetilde{\Delta}_2$ and $\textrm{Re}\, \widetilde{\Delta}_3$. 
These formulae for the potentials ${\rm Re} ~\widetilde{\omega}_i$, ${\rm Re} ~\widetilde{\Delta}_I$ in the microscopic theory all agree precisely with the corresponding gravitational formulae for $\partial_T\Phi_I$ and $\partial_T\Omega_i$ given in (\ref{eq:PhiadT}-\ref{eq:dOmegaadT}). This confirms the identifications \eqref{eqn:reDelderT}. 

It is significant that these comparisons {\it all} agree. The differentials $(d{\rm Re} ~\widetilde{\omega}_i, d{\rm Re} ~\widetilde{\Delta}_I)$
form a vector in the five dimensional space that is generated (locally) by the direct sum of the tangent space to the BPS surface and its normal, and the latter violates the constraint. 
It is also noteworthy that the comparisons agree {\it precisely}. Formulae in the BPS limit are only defined modulo the constraint on charges but the agreements found here
apply also without using the constraint. These facts suggest that the microscopic description goes beyond BPS. 

We can make these comments quantitative by forming the differential 
\begin{equation}
\label{eq:MicroFirstLawRealPart}
dS^*  + {\rm Re}\, \widetilde{\Delta}_I\, dQ_I + {\rm Re}\, \widetilde{\omega}_i\, dJ_i\, 
= \frac{2 \pi^2}{S^*} \frac{(Q_1 + Q_2 + Q_3 + \frac{1}{2} N^2)\, dh}{ \left(\frac{S^*}{2\pi}\right)^2 + (Q_1 + Q_2 + Q_3 + \frac{1}{2} N^2)}\, ,
\end{equation}
where the height function $h$ defined through 
\be\label{eq:ChargeConstrain}
 h = \Bigg(Q_1 Q_2 Q_3 + \frac{1}{2} N^2 J_1 J_2\Bigg)  - \Bigg(Q_1 + Q_2 + Q_3 + \frac{1}{2} N^2\Bigg)  \Bigg(Q_1 Q_2 + Q_2 Q_3 + Q_3 Q_1 - \frac{1}{2} N^2(J_1 + J_2) \Bigg) \, ,
\ee
is a measure of the violation of the constraint \eqref{eq:ChargeConstraint} on black hole charges. The addition of the differential $dS^*$ on the left hand side of \eqref{eq:MicroFirstLawRealPart} removes the terms that are attributable to BPS physics. We can interpret the remainder as a formula for the entropy in excess of the BPS
entropy
\be\label{eq:dS-dS*}
S - S^* =  \frac{2 \pi^2 h}{S^*} \frac{Q_1 + Q_2 + Q_3 + \frac{1}{2} N^2\, }{ \left(\frac{S^*}{2\pi}\right)^2 + (Q_1 + Q_2 + Q_3 + \frac{1}{2} N^2)} \, , 
\ee
due to the violation of the constraint $h=0$. This formula agrees with the gravitational result \eqref{eqn:dhinDirstLaw}. We point out (again) that our definition \eqref{eq:ChargeConstrain}
of the height function is such that $h\geq 0$ corresponds to positive entropy. 

It is also well worth the effort to extract the imaginary parts of the complex potentials (\ref{eq:SolOmegai}-\ref{eq:SolDeltaI}) found in the microscopic computation. Representative components are
\begin{align}
  \textrm{Im}\, \widetilde{\omega}_1 \over 2 \pi & = - \frac{1}{4}N^2 \frac{J_2 +Q_1 + Q_2 + Q_3 + \frac{1}{2} N^2}{\left(\frac{S^*}{2 \pi} \right)^2 + (Q_1 + Q_2 + Q_3 + \frac{1}{2} N^2)^2}\, ,\label{eq:ImOmegai}\\
  \textrm{Im}\, \widetilde{\Delta}_1 \over 2 \pi & = \frac{(Q_2 + Q_3) (2Q_1 + Q_2 + Q_3 + \frac{1}{2} N^2)-\frac{1}{2} N^2(J_1 + J_2) }{2 \Big[\left(\frac{S^*}{2 \pi} \right)^2 + (Q_1 + Q_2 + Q_3 + \frac{1}{2} N^2)^2\Big]}\, ,\label{eq:ImDelta1}
\end{align}
and the complete set of potentials follow by appropriate permutation of indices. 
It is instructive to collect the entire vector of imaginary potentials as a one-form
\be\label{eq:ImPotOneform}
 {\textrm{Im}\, \widetilde{\Delta}_I  dQ_I  + \textrm{Im}\, \widetilde{\omega}_i dJ_i \over 2\pi}= - \frac{dh}{2[\left(\frac{S^*}{2 \pi} \right)^2 + (Q_1 + Q_2 + Q_3 + \frac{1}{2} N^2)^2]}~.
\ee
The entropy function formalism for the BPS black holes in AdS$_5$ \cite{Hosseini:2017mds} employs complex potentials and identify their real part as the physical potential, but 
their imaginary part is enigmatic. It is therefore satisfying that the imaginary potentials computed in the microscopic description of BPS black holes vanish on the constraint surface $h=0$ 
but increase proportionally to violations of the constraint.  

Combining the real and imaginary parts of the potentials (\ref{eq:MicroFirstLawRealPart}, \ref{eq:ImPotOneform}), we can write the first law of thermodynamics as a complex-valued expression:
\begin{align}
  {} & T dS^* + (\widetilde{\Delta}_I - \widetilde{\Delta}_I^*) dQ_I + (\widetilde{\omega}_i - \widetilde{\omega}_i^*) dJ_i \nonumber\\
  = &   \, \left(\frac{2 \pi}{S^*} (Q_1 + Q_2 + Q_3 + \frac{1}{2} N^2) T - i \right) \frac{\pi\, dh}{\left(\frac{S^*}{2 \pi} \right)^2 + (Q_1 + Q_2 + Q_3 + \frac{1}{2} N^2)^2}\, .\label{eq:CplxFirstLaw}
\end{align}
It is not a coincidence that this formula is a close analogue of \eqref{eqn:dhinDirstLaw}, the consolidated formula for all the potentials we computed from gravity in the nearBPS region. In the following subsection we relate them quantitatively. 

\subsection{Extremizing Free Energy nearBPS}
\label{sec:nearBPSext}

In this subsection, we generalize the BPS entropy extremization reviewed in subsection \ref{sec:Extremization} to an extremization principle that describes the nearBPS region. 

\subsubsection{The Free Energy Function}
As we stressed repeatedly in section \ref{sec:FreeFT}, all microscopic computations of the BPS 
arrive at the partition function \eqref{eqn:BPSparti} subject to the subsidiary condition \eqref{eq:BPSconstr}
and the black hole entropy is subsequently extracted
therefrom through the extremization procedure reviewed in subsection \ref{sec:Extremization}. The variables of the BPS partition function are potentials that must be interpreted as thermal derivatives of physical potentials. Our discussion in subsection \ref{sec:firstlawmicro} arrived at this identification in the course of comparing microscopic conventions with those of gravity, but it applies equally as a general relation between the extremal and near-extremal partition functions. Thus the adaptation of the BPS partition function \eqref{eqn:BPSparti} to a notation appropriate for the nearBPS regime is 
\be
\label{eqn:logZextnear}
  \ln~Z (\beta, \Delta_I, \omega_i) = - \frac{N^2}{2 T} \frac{(\Delta_1 - \Delta_1^*) (\Delta_2 - \Delta_2^*) (\Delta_3 - \Delta_3^*)}{(\omega_1 - \omega_1^*) (\omega_2 - \omega_2^*)}\, .
\ee
It is significant to note that there are {\it no} tilde's on the variables in this formula: $\Delta_I$ and $\omega_i$ refer to the full potentials rather than their thermal derivatives. The BPS reference values
$\Delta^*_I$ and $\omega^*_i$ all equal $1$ numerically, because the BPS mass is the sum of the conjugate conserved charges with coefficient $1$ (in units where $\ell_5=1$). We maintain the more elaborate notation for conceptual clarity. The original BPS partition function \eqref{eqn:BPSparti} is recovered in the extremal limit $T\to 0$, as it should be.  

The microscopic partition function \eqref{eqn:physZtar} was introduced generally, without restricting to the BPS (or even the nearBPS) regime. In the saddle point approximation 
it gives
\be
\label{eqn:saddlelnZ}
  \ln ~Z  = S  - \beta(M-M^*)  + \beta(\Delta_I - \Delta_I^*)Q_I  + \beta(\omega_i - \omega_i^*)J_i ~.
\ee
It is tempting to identify $\ln Z$ in this formula with the BPS partition function \eqref{eqn:logZextnear} but that is incorrect even in the strict BPS limit where it is crucial that we require the potentials to
satisfy the complex constraint \eqref{eq:BPSconstr}. Moreover, in the BPS limit the real part of the potentials satisfy $\sum_I(\Delta_I - \Delta_I^*)-\sum_i(\omega_i - \omega_i^*)=0$
by definition but this equation must be relaxed in the nearBPS region. We impose the constraint
\be
\label{eqn:defconstra}
\sum_I (\Delta_I - \Delta_I^*) - \sum_i (\omega_i - \omega_i^*)  = \varphi + 2 \pi i T ~.
\ee
The imaginary part of this equation reformulates the BPS condition \eqref{eq:BPSconstr} in a manner that is meaningful also for small but nonvanishing temperature. The real part
allows potentials to depart from their BPS values and parametrize this violation by $\varphi$, in conformity with the convention in gravity \eqref{eqn:varphidef}.

We now combine the general saddle point approximation \eqref{eqn:saddlelnZ} with the microscopic computation of the free energy \eqref{eqn:logZextnear} and subject the result to the complex constraint \eqref{eqn:defconstra}. This gives the free energy in the nearBPS regime
\begin{align}
\label{eqn:nearBPSfree}
  \mathcal{F} & \equiv (M - M^*) - TS \nonumber\\
  {} & = \frac{1}{2} N^2 \frac{(\Delta_1 - \Delta_1^*) (\Delta_2 - \Delta_2^*) (\Delta_3 - \Delta_3^*)}{(\omega_1 - \omega_1^*) (\omega_2 - \omega_2^*)}
  +(\Delta_I - \Delta_I^*)Q_I  + (\omega_i - \omega_i^*)J_i +
  \nonumber\\
  {} & \quad + \Lambda \Big( \sum_I (\Delta_I - \Delta_I^*) - \sum_i (\omega_i - \omega_i^*) - \varphi - 2 \pi i T \Big)\, .
\end{align}
We have not attempted to argue that there can be no additional contributions to the free energy in the nearBPS regime. On the contrary, we conservatively 
claim that it {\it at least} includes the ingredients incorporated in
\eqref{eqn:nearBPSfree}.  

An {\it effective} theory of nearBPS black holes can be found be extremizing the free energy \eqref{eqn:nearBPSfree} over the potentials $\Delta_I,\omega_i$, and the Lagrange multiplier $\Lambda$. 
It will depend on the remaining potentials $\varphi, T$, the dynamical fields in the effective description. As usual, the effective theory will also feature dimensionful parameters that are unspecified {\it a priori} but computable from the UV completion in principle. In the present context a complete microscopic theory relates the effective parameters ($C_T$, $C_\varphi$, $C_E$) to
conserved charges. 
 
 \subsubsection{Extremization of Free Energy}
 
The free energy \eqref{eqn:nearBPSfree} differs from the entropy function \eqref{eq:EntropyFct} only by an overall factor $-T$ and the addition of a simple term $-\Lambda \varphi$. Therefore the extremization is nearly unchanged from the BPS computation in subsection \ref{sec:Extremization}. 
The equations of motion (\ref{eq:BPSExtEq-1}-\ref{eq:BPSExtEq-2}) are entirely unchanged so the ratios (\ref{eqn:Delomratio}-\ref{eqn:omegaratio}) remain, and so the steps 
needed for finding the potentials explicitly are exactly the same. The key modification is the equation of motion for $\Lambda$, ie. the constraint \eqref{eqn:defconstra}. For the potentials the constraint enters in \eqref{eqn:keyom} where its role is to provide an overall normalization for the potentials that are otherwise determined by relations between their ratios. 
The new (complex) normalization modifies the potentials from (\ref{eq:SolOmegai} - \ref{eq:SolDeltaI}) to
\begin{align}
  \frac{\omega_i - \omega^*_i}{\varphi + 2 \pi i T} & = \frac{1}{2} N^2\frac{\prod_k (J_k - \Lambda)}{J_i - \Lambda} \frac{1}{2 \Lambda (\Lambda + Q_1 + Q_2 + Q_3 + \frac{1}{2} N^2)}\, ,\\
  \frac{\Delta_I - \Delta_I^*}{\varphi+ 2 \pi i T} & = \frac{\prod_K (Q_K + \Lambda)}{Q_I + \Lambda} \frac{1}{2 \Lambda (\Lambda + Q_1 + Q_2 + Q_3 + \frac{1}{2} N^2)}\, .
\end{align}
Consider $\omega_1$ for definiteness. 
For small $\varphi$ the constraint on the charges is only violated mildly so we assume that $\Lambda$ is unchanged to leading order, and then the right hand of the first equation 
is ${1\over 2\pi i } \left( {\rm Re} ~\widetilde{\omega}_1 + i {\rm Im}~\widetilde{\omega}_1\right) $ where  ${\rm Re} ~\widetilde{\omega}_1$ and ${\rm Im}~\widetilde{\omega}_1$ are
the BPS values for these variables given in (\ref{eq:ReOmegai}, \ref{eq:ImOmegai}). The physical potential is the real part so the leading dependence on the constraint violation $\varphi$
is due to the {\it imaginary} part of $\widetilde{\omega}_1$:
\be
\label{impottovarphi}
\left.{\rm Re} (\omega_1 - \omega^*_1) \right|_{\varphi~{\rm dependence}} = \varphi~ {{\rm Im}~\widetilde{\omega}_1 \over 2\pi} 
= - \frac{1}{4}N^2 \frac{J_2 +Q_1 + Q_2 + Q_3 + \frac{1}{2} N^2}{\left(\frac{S^*}{2 \pi} \right)^2 + (Q_1 + Q_2 + Q_3 + \frac{1}{2} N^2)^2} \varphi  ~.
\ee
This agrees exactly with the gravitational computation of the change of the rotational velocity $\Omega_1$ due to 
small violations of the constraint \eqref{eq:dOmegaadvarphi}. The dependences of all other potentials on $\varphi$ similarly agree with gravitational results. 

Hitherto, we have showed that the extremization principle based on the free energy \eqref{eqn:nearBPSfree} reproduces the dependence of all 
physical potentials on temperature and on violations of the constraint parametrized by the field deformation $\varphi$. We now turn to to the mass and entropy in the nearBPS regime. 

As we mentioned earlier in this subsection, the equations of motion are entirely unchanged from the BPS case (\ref{eq:BPSExtEq-1}-\ref{eq:BPSExtEq-2}), except for the 
equation of motion for the Lagrange multiplier $\Lambda$, ie. the constraint \eqref{eqn:defconstra}. Therefore the cubic equation \eqref{eq:CubicLambda} for $\Lambda$ still holds.
However, since the potentials $\Delta_I, \omega_i$ do not satisfy the BPS constraint \eqref{eq:BPSconstr} in the nearBPS theory, the charges $Q_I, J_i$ may also violate their constraint \eqref{eq:ChargeConstraint}. 

We already anticipated this situation in the preceding subsection when introducing a height function $h$ \eqref{eq:ChargeConstrain} parametrizing the violation of the constraint \eqref{eq:ChargeConstraint}. In the schematic notation introduced in (\ref{eq:CubicExplicit-1}-\ref{eqn:ADCdef}) the height function $h \equiv C - A B$ deforms the cubic equation
to
\be\label{eq:CubicEq-2}
\Lambda^3 + A\Lambda^2 + B\Lambda + C =   (\Lambda^2 + B) (\Lambda + A)  + h = 0 \, .
\ee
Modifying charges and $\Lambda$ away from $h=0$ at first order in perturbation theory 
$$
(2\Lambda \delta \Lambda + \delta B)(\Lambda + A) + h = 0~,
$$
and recalling that $\Lambda$ is purely imaginary at leading order we find
\be 
\label{eqn:ReImLam}
{\rm Im} ~\delta \Lambda =    { A h\over 2{\rm Im} \Lambda (B+A^2)}  + {\delta B\over 2{\rm Im}\Lambda}  ~.
\ee
The second term 
\be
{\delta B\over 2{\rm Im}\Lambda}  = - {\delta S_*\over 2\pi}~, 
\ee
takes into account the change of the BPS entropy due to changes of the conserved charges. Therefore
the entropy in excess of the BPS entropy due to the violation of the constraint becomes
\be
\delta (S-S^*) =  - 2\pi {\rm Im} ~\delta \Lambda =  {2\pi^2 h\over S^*} { Q_1 + Q_2 + Q_3 + \frac{1}{2}N^2\over \left({S^*\over 2\pi}\right)^2  + (Q_1 + Q_2 + Q_3 + \frac{1}{2}N^2)^2 } ~.
\ee
This formula agrees with the expression (\ref{eq:dS-dS*}) inferred from applying the thermodynamic potentials derived in the BPS case also off the BPS surface. 
Moreover, the same expression was found also from gravity considerations \eqref{eqn:dhinDirstLaw}. 
The considerations here are based on a minimal generalization of the {\it microscopic} theory.

\subsubsection{Parameters in Effective Field Theory}

We have expressed possible violations of the constraint in two distinct ways. We defined the height 
function $h$ \eqref{eq:ChargeConstrain} that measures departures from the constraint $h=0$ on charges, and the effective potential $\varphi$ introduced through the deformed constraint \eqref{eqn:defconstra} that is required to take the value $\varphi=0$ by the conditions for 
supersymmetry \eqref{eq:BPSconstr}. The ``charge" $h$ and the ``potential" $\varphi$ are proportional near the BPS surface we computed their constant of proportionality \eqref{eqn:hvsvarphi} on the gravity side. It is interesting to understand their relation more generally. 

The height function $h$ \eqref{eq:ChargeConstrain} is fairly elaborate so the expression for small departures $\delta h = \partial_{Q_I}h\delta Q_I + \partial_{J_i}h\delta J_i$ from the constraint $h=0$ due to general variations $\delta Q_I$, $\delta \omega_i$ of the charges is lengthy and not illuminating. 
However, variations that are {\it proportional} to the charges themselves  $\delta Q_I=\lambda Q_I$, $\delta J_i=\lambda J_i$ yield a manageable formula 
\be
 \delta h = - \Big[\left(Q_1^2 (Q_2 + Q_3) + \textrm{cyclic} \right) + 2 Q_1 Q_2 Q_3 + \frac{1}{4} N^4 (J_1 + J_2) \Big] \lambda~,
\ee
after simplifications using the constraint $h=0$. However, the transformation of the effective potential is canonical (with dimension $-2$ relative to the charge):
\be
\delta \varphi = - 2\lambda \varphi~.
\ee
Comparison of the two preceding equations gives
\be
\label{eqn:handlamrel}
 h = {1\over 2} \Big[\left(Q_1^2 (Q_2 + Q_3) + \textrm{cyclic} \right) + 2 Q_1 Q_2 Q_3 + \frac{1}{4} N^4 (J_1 + J_2) \Big] \varphi~.
\ee
This generalizes the constant of proportionality \eqref{eqn:hvsvarphi} computed in the gravity to the case of three distinct charges. 

We can now consolidate our results by presenting the first law of thermodynamics in a coherent manner. 
We collected most of them already in the complex form of the first law \eqref{eq:CplxFirstLaw}.
In the course of this subsection we have rederived each of the terms in this equation from free energy extremization. Additionally, the 
imaginary part of the potentials acquired a more satisfying interpretation through its relation with the deformation parameter $\varphi$ given in 
\eqref{impottovarphi} and its analogues for other potentials. The conversion \eqref{eqn:handlamrel} between $h$ with $\varphi$ finally let 
us rewrite \eqref{eq:CplxFirstLaw} as
\be
  T dS^* + (\Delta_I - \Delta_I^*) dQ_I + (\omega_i - \omega_i^*) dJ_i  = \frac{C_T}{T} \Bigg[\frac{2 \pi}{S^*} (Q_1 + Q_2 + Q_3 + \frac{1}{2} N^2) T + \frac{\varphi}{2 \pi} \Bigg] \frac{d\varphi}{2 \pi} \, ,
\ee
where $C_T$ is the heat capacity that is linear in temperature with proportionality constant
\begin{eqnarray}
  \frac{C_{T}}{T}\bigg|_{\text{nearExt}}         
         &=&\pi^2 \frac{Q_1^2 (Q_2 + Q_3) +Q_2^2 (Q_3 + Q_1) +Q_3^2 (Q_1 + Q_2)+ 2 Q_1 Q_2 Q_3 + \frac{1}{4} N^4 (J_1 + J_2)}{Q_1Q_2 +Q_2 Q_3 + Q_3 Q_1-\frac{1}{2} N^2 \left(J_1 + J_2\right) + (Q_1 + Q_2 + Q_3 + \frac{1}{2}N^2)^2}~.\nonumber
\end{eqnarray}
We previously computed this expression in gravity \eqref{heat capacity in terms of charges}. In some parts of this paper we introduced the effective field theory 
parameters $C_\varphi$ and $C_E$ in addition to $C_T$. They all have different physical significance but numerically $C_\varphi=C_T$  and
\be
\left({C_E\over T}\right) = {2\pi\over S^*} \left({C_T\over T}\right) (Q_1 + Q_2 + Q_3 + \frac{1}{2} N^2)~,
\ee
so in the presentation of these final results we opt for writing the formulae more explicitly. 
           
The first law of thermodynamics
\be
T dS = d(M - M^*) - (\Delta_I - \Delta_I^*) dQ_I - (\omega_i - \omega_i^*) dJ_i ~,
\ee
now gives the energy and the entropy of the excitations above the BPS ground state as
\begin{align}
  M - M^* & = \frac{1}{2} \frac{C_T}{T} \left(\frac{\varphi}{2 \pi} \right)^2\, ,\\
  S - S^* & = \frac{2 \pi}{S^*} \frac{C_T}{T} (Q_1 + Q_2 + Q_3 + \frac{1}{2} N^2) \frac{\varphi}{2 \pi}\, .
\end{align}
These expressions agree with results earlier in the paper, including computations in gravity (\ref{eq:ExtNearBPSM-M*}, \ref{eqn:SmSstar}). 

Our microscopic discussion in this section has not at all touched on the conventional heat capacity, ie. the term in the entropy that is linear in temperature (and correlated with a mass term that is quadratic in the temperature). What is needed to get this term is an equation of motion where the height parameter $h$ in the cubic equation \eqref{eq:CubicEq-2} is traded for $\varphi$ through \eqref{eqn:handlamrel} and subsequently complexified to $\varphi + 2\pi i T$. The last step is very natural in that much of the theory apparently depends holomorphically on a 
complex symmetry breaking parameter. However, we are not yet able to present a principled argument based on microscopic theory.

\section{Summary and Outlook}\label{sec:discussion}

In this paper we have discussed the thermodynamics and the statistical physics of AdS$_5$ black holes, addressing both the BPS and the nearBPS configurations. 
In the nearBPS region, we made an important distinction between the near-extremal ($T\neq 0$, $\varphi=0$), the extremal nearBPS ($T = 0$, $\varphi\neq 0$), and the general 
nearBPS  ($T \neq 0$, $\varphi\neq 0$) black holes. The unfamiliar potential $\varphi$ parametrizes the possible violation of a constraint on charges that must be imposed in the strict BPS limit. 

In the gravitational theory we studied all thermodynamic potentials in great detail, especially their interrelation through the first law of thermodynamics, in the entire BPS and nearBPS domain. 
We also reviewed an elementary version of the holographically dual microscopic description, based on the free field representation of ${\cal N}=4$ SYM. It yields the semiclassical 
partition function found in \cite{Hosseini:2017mds} that is common to all the recent proposals for a microscopic theory of BPS black holes in 
AdS$_5$~\cite{Cabo-Bizet:2018ehj, Choi:2018hmj, Benini:2018ywd}. 

We found that, with minor additional assumptions, the same semiclassical partition function also describes aspects of nearBPS black holes. This approach to theory without supersymmetry is particularly successful for extremal nonBPS black holes. It is the main basis for our claim of a microscopic description of the nearBPS black holes. Our generalization of the entropy function to a free energy function captures the effective field theory of the nearBPS region succinctly and in a manner that relates 
directly to the microscopic theory. 

The recent progress towards a statistical description of BPS black holes~\cite{Cabo-Bizet:2018ehj, Choi:2018hmj, Benini:2018ywd} is not yet fully satisfactory. It is not even clear that the reports 
are consistent with one another. Our study gives general support for these advances. For example, we find that the BPS limit is robust in that it can be approached from any direction. 

However, there are still many open questions, particularly on the microscopic side. Our incorporation of mass is tenuous, especially for the near-extremal black holes. 
The free energy function could surely be improved. The more important open problem is why the microscopic description of nearBPS black holes is even possible. 
It appears that some non-renormalization due to supersymmetry persists also away from the BPS limit. 

Furthermore, we expect that the additional microscopic degrees of freedom in the nearBPS region can equally be modeled by a simple gas of free particles, much like the free model of the 
BPS limit that we review. However, it might well be necessary to invoke other sectors of ${\cal N}=4$ SYM that are BPS, but preserving different supersymmetries than the ground state. 
This is the structure of successful microscopic models for near-extremal black holes with flat asymptotic space, such as the D1-D5 system \cite{Callan:1996dv}.

There are also problems in gravitational physics that we leave for the future. We found that the potential for constraint violations $\varphi$ exhibits entropic preference for $\varphi\geq 0$, reminiscent of the behavior of conventional temperature $T$. It would be interesting to develop the geometric underpinnings of $\varphi$. We anticipate that nonnegative $\varphi$ is imposed by the absence of closed time-like curves. Additionally, the near horizon AdS$_2$ expected for any near-extremal black hole must have an analogue for extremal nearBPS configurations and the full nearBPS region of parameter space promises an interesting interplay between the low temperature limit and a mildly violated constraint. 

We look forward to pursue these and related directions in future research.  

\section*{Acknowledgements}

We would like to thank Francesco Benini, Seyed Hosseini, Joonho Kim, Dario Martelli, Sameer Murthy, Vasily Pestun, Elli Pomoni, Wei Song and Yang Zhou for many helpful discussions and communications.  This work was supported in part by the U.S. Department of Energy under grant DE-FG02-95ER40899. FL would like to thank UCLA, UCSD, ICTP, IPMU and YITP for hospitality. A preliminary version of this work was presented at the YITP workshop ``Quantum Information and String Theory 2019'' on May 30, 2019. JN's work was supported in part by the U.S. Department of Energy under grant DE-SC0007859 and by a Van Loo Postdoctoral Fellowship.

\providecommand{\href}[2]{#2}\begingroup\raggedright\endgroup

\end{document}